\documentclass[useAMS,usenatbib]{mn2e} 
\usepackage[totalwidth=500pt,totalheight=700pt]{geometry}
\usepackage{natbib}

\usepackage{mathptmx} 
\usepackage{sdssp1}
\usepackage{graphicx}
\usepackage{amssymb}
\usepackage{ctable}

\title[Star formation in SDSS close pairs]{Star Formation in Close Pairs Selected from the Sloan
  Digital Sky Survey}

\author[B. Nikolic et al.]{B.~Nikolic\thanks{Supported by a PPARC Studentship}
  H.~Cullen\thanks{Supported by a PPARC Studentship} P.~Alexander \\
Astrophysics Group, Cavendish Lab., Cambridge CB3 0HE}

\date{2004/03/14 }
\pagerange{\pageref{firstpage}--\pageref{lastpage}; } \pubyear{2004}

\begin{document}

\label{firstpage}
\maketitle 

\begin{abstract}

The effect of galaxy interactions on star formation has been
investigated using Data Release 1 of the Sloan Digital Sky Survey
(SDSS). Both the imaging and spectroscopy data products have been used
to construct a catalogue of nearest companions to a volume limited
($0.03 < z < 0.1$) sample of galaxies drawn from the main galaxy
sample of SDSS. Of the 13973 galaxies in the volume limited sample, we
have identified 12492 systems with companions at projected separations
less than 300\,kpc.  Star-formation rates for the volume-limited
sample have been calculated from extinction and aperture corrected
H$\alpha$ luminosities and, where available, \emph{IRAS}
data. Specific star formation rates were calculated by estimating
galaxy masses from $z$-band luminosities, and $r$-band concentration
indices were used as an indicator of morphological class.  The mean
specific star-formation rate is significantly enhanced for projected
separations less than 30\,kpc. For late-type galaxies the correlation
extends out to projected separations of 300 kpc and is most pronounced
in actively star-forming systems.  The specific star-formation rate is
observed to decrease with increasing recessional velocity difference,
but the magnitude of this effect is small compared to that associated
with the projected separation.  We also observe a tight relationship
between the concentration index and pair separation; the mean
concentration index is largest for pairs with separations of
approximately 75\,kpc and declines rapidly for separations smaller
than this. This is interpreted as being due to the presence of
tidally-triggered nuclear starbursts in close pairs.  Further, we find
no dependence of star formation enhancement on the morphological type
or mass of the companion galaxy.

\end{abstract}


\section{Introduction}

It is well established, both observationally and theoretically, that
galaxies do not evolve in isolation from one another but that their
present day structure is the result of sequential mergers and
encounters of galaxies and proto-galaxies. 
For example, \cite{1972ApJ...178..623T}
used numerical simulations to show that many of the features of
peculiar galaxies can be explained by recent galaxy-galaxy
interactions and mergers. Indeed, in the currently widely favoured
hierarchical models of galaxy formation, all of the galaxies at the
present epoch are the results of mergers and accretion
\citep[e.g.,][]{2000MNRAS.319..168C}.

The problem of predicting the outcomes of galaxy encounters is
difficult. Much progress has been made using numerical simulations
\citep[e.g.,][]{1996ApJ...464..641M}, following the pioneering work of
\cite{1972ApJ...178..623T}, but several outstanding problems
remain. The fate of the inter-stellar medium (ISM) and eventual
conversion to new stars is particularly difficult to follow due to the
finite resolution of simulations and the complicated physics of
the ISM. In this paper we investigate one aspect of this problem, the
effect of tidal interaction on the star formation rate of galaxies,
using data from the Sloan Digital Sky Survey
\citep[SDSS;][]{2000AJ....120.1579Y}.

The connection between star formation and galaxy interactions has been
suspected ever since the first large surveys, biased towards star
forming systems were conducted: first in the ultraviolet continuum,
then in emission lines and the far infrared. The far infrared survey
by the Infrared Astronomical Satellite ({\it IRAS}), in particular,
found a population of star forming systems in the nearby universe with
$L_{\rm FIR} > 1\times10^{11}L_{\odot}$, corresponding to an inferred
star formation rate greater than $22\,M_{\odot}$\,year$^{-1}$. These
star-forming galaxies exhibit, on average, more peculiar morphologies
and are more likely to be mergers. Furthermore, these observational
indications of interaction/merger become more common and pronounced in
systems with higher star-formation rates. The most luminous of the
{\it IRAS\/} sources, i.e., those with $L_{\rm FIR}\geq
3\times10^{12}\,L_{\odot}$, being universally classified as strong
interactions or mergers \citep{1996ARA&A..34..749S}.

A number of existing studies have used samples of interacting galaxies
to examine statistically the causal relationship between galaxy
interactions and star formation. For example,
\cite{1978ApJ...219...46L} found a much higher dispersion in the $ U-
B$ versus $ B- V$ colour-colour plot for galaxies selected from Arp's
atlas (1966) as compared to a control sample, indicating an enhanced
star formation rate within the last $\sim 10^8$
years. \cite{1987AJ.....93.1011K} used H$\alpha$ line and far infrared
observations to examine the influence of interactions on the global
star formation of a complete sample of close pairs, and a sample of
Arp systems.  They found a higher than average star formation rate in
interacting systems, but also found that a large fraction of galaxies
in close pairs exhibit normal star formation rates. Similarly,
\cite{1988ApJ...335...74B} examined the far infrared emission in a
sample of strongly interacting galaxy pairs, finding enhanced emission
in some, but not all systems. \cite{2003A&A...405...31B} have looked
at star formation in a sample of 59 interacting and merging systems,
and 38 isolated galaxies, using spectroscopic and photometric
observations in the optical/near-infrared.  In contrast to other
results, they find that the global $UBV$ colours do not support
significant enhancement of the star-formation rate in
interacting/merging galaxies.

Defining samples of interacting or merging galaxies to investigate the
effects of interaction on star formation is not straight forward. One
technique, based on the work of \cite{1972ApJ...178..623T}, is to
select galaxies with peculiar morphologies, especially those showing
tidal tails or galactic bridges. Such systems are often drawn from
Arp's Atlas of Peculiar Galaxies
(\citeyear{1966apg..book.....A}). This technique has been used by a
number of authors, including, \cite{1978ApJ...219...46L} and
\cite{1987AJ.....93.1011K}. The drawback, as illustrated by
\cite{1972ApJ...178..623T} themselves, is that the degree of
morphological disturbance is sensitive to the orbital parameters of
the interaction, and so can not be relied upon to trace all
interactions. Additionally, the features characteristic of interaction
are often of low surface brightness, requiring deep imaging for
detection, and must be identified by visual inspection.

When the interacting systems are separated by distances comparable to,
or larger than, their optical extents, it is relatively easy to
resolve the galaxies in imaging data and, based on their projected
separation, identify them as a physical pair. If spectral data are
available, the velocity separation can also be used. Using this
technique, survey data can by analysed for close pairs yielding large
samples of interacting galaxies, albeit usually missing the galaxies
in the later stages of interaction and merger.

Large area redshift surveys have enabled studies of large samples of
interacting galaxies. For example, \cite{ 2000ApJ...530..660B} have
analysed optical spectra from a sample of 502 galaxies in close pairs
and N-tuples from the CfA2 redshift survey finding the equivalent
widths of H$\alpha$ anti-correlate strongly with pair spatial and
velocity separation. Similarly, \cite{2003MNRAS.346.1189L} examined
1853 pairs in the 100k public release of the 2dF galaxy survey and
find star formation in galaxy pairs to be significantly enhanced over
that of isolated galaxies for separations less than 36\,kpc and
velocity differences less than $100\,\kms$. 

In this paper we follow a similar approach and use the SDSS to define
a large sample of 12861 galaxies with identified companions. Our aim
is to establish the relative importance of galaxy interactions in
determining star formation. 

A number of studies have established a link between density of
environment and star-formation rate, for example using the SDSS
\citep{2003ApJ...584..210G}, 2dF \citep{2002MNRAS.334..673L}, and Las
Campanas Redshift Survey \citep{1998ApJ...499..589H}. In all three
studies a correlation is found between density of environment and
star-formation rate in the sense that ongoing star formation is
suppressed in the denser environments.  \cite{2003ApJ...584..210G} and
\cite{1998ApJ...499..589H} have attempted to decouple this effect from
the density-morphology \citep{1980ApJ...236..351D} or radius
morphology relations \citep{1993ApJ...407..489W} finding that the
density/star-formation rate relationship exists independent of the
density-morphology relationship.

The structure of this paper is as follows: In Section 2 we discuss the
compilation of a volume- and luminosity-limited sample drawn from the
SDSS, the method used to define galaxy pairs together with details of
various physical parameters calculated for each system. In Section 3
we present our results and in Section 4 we discuss these results in
the context of earlier work and consider the implications for our
understanding of interaction-triggered star formation.

\section{The Data}
\subsection{Sloan Digital Sky Survey}
\label{sec:sloan-digital-sky}

The Sloan Digital Sky Survey (SDSS) is an imaging survey in five broad
photometric bands \citep[{\it u, g, r, i,} and {\it z\/}, defined
in][]{1996AJ....111.1748F} with medium-resolution ($R\approx
1800$--$2100$) spectroscopic follow up of approximately one million
targets. When complete, it will cover most of the Northern Galactic
Hemisphere. The sample used in this work is based on Data Release 1
(DR1) which covers 1360\,deg$^{2}$ of the sky; however, we use the
improved spectroscopic data for each object from Data Release 2 (DR2).

The Main Galaxy Sample \citep[MGS,][]{2002AJ....124.1810S} was
selected from the SDSS imaging data as a galaxy sample for
spectroscopic observations. Objects were identified as galaxies in the
SDSS imaging data if their \rband\ magnitude measured using the
best-fitting galaxy light curve was at least 0.3 magnitudes brighter
the magnitude measured using the point spread function model. All such
galaxies with \rband\ Petrosian magnitudes brighter than 17.77 were
selected for spectroscopic follow up and were observed, as far as
possible, given the fibre-placement constraints and the finite number
of fibres on any one plate. The resulting completeness is 93\%. Of the
7\% of galaxies that are missing, 6\% are galaxies which could not be
observed due to the minimum fibre separation constraint.

The MGS was used to construct (Section~\ref{sec:primary-catalogue}) a
volume limited sub-sample. Each member of this sub-sample was than
paired with its nearest galaxy (Section~\ref{sec:defin-catal-galaxy})
and its star formation rate estimated using two methods
(Section~\ref{sec:star-formation-rate}).

\subsection{Primary Catalogue}
\label{sec:primary-catalogue}

Our primary catalogue is a complete, volume- and luminosity-limited,
sub-sample of the MGS.  It was derived from the MGS using a procedure
analogous to that used by \cite{2003ApJ...584..210G}. The volume of
the sample was defined by the redshift constraint $0.03<z<0.1$, where
the lower limit was used to avoid subsequent large aperture
corrections when estimating star-formation rates from \Halpha
measurements (Section \ref{sec:halpha-line}). The upper redshift limit
allows us to construct a complete sample by retaining only galaxies
with \rband\ absolute magnitudes $M_{r} < -20.45$.

Two further criteria were needed to remove a small number of spurious
entries. The first was the elimination of galaxies with redshift
confidence less than 0.7. This relatively relaxed constraint
removes most of the galaxies with poorly determined redshifts while
introducing only a very small bias towards active emission lines
systems.  Secondly, entries in the SDSS with Petrosian \zband\
magnitudes fainter than 22.83, i.e., below the detection limit
\citep[Table 21 in ][]{2002AJ....123..485S}, were excluded. Since
galaxies in the main sample have $r < 17.77$, entries with $z> 22.83$
have extreme colours and are almost certainly spurious detections.

For the purposes of this study, we wish to remove systems in which an
AGN provides the dominant contribution to the ionising radiation (in
these cases an accurate estimate of the star-formation rate is not
possible using the methods we employ).  Following
\cite{1987ApJS...63..295V}, AGN were identified by their positions on
the $[\hbox{N{\sc\,ii}}]/\Halpha$ vs $[\hbox{O{\sc\,iii}}]/\Hbeta$
and, in a few cases, $[\hbox{O{\sc\,i}}]/\Halpha$ vs
$[\hbox{O{\sc\,iii}}]/\Hbeta$ planes. We required the signal to noise
ratio for the line fluxes to exceeded 2 before taking them into
consideration.  Approximately 38\% of galaxies from the volume limited
sample were identified as having an AGN.  Of these 64\% are early- or
mixed-type galaxies. A further 295 objects which were not classified
spectroscopically as AGN but exhibited very broad \Halpha lines with
$\sigma > 570\,\kms$, were also excluded.

\subsection{Star Formation Rate Estimators}
\label{sec:star-formation-rate}

\cite{1998ARA&A..36..189K} reviews a number of widely used estimates
for galaxy star-formation rate (SFR), including H$\alpha$ and
$[$O{\sc\,ii}$]$ emission lines, together with ultraviolet,
far-infrared (FIR) and radio continuum emissions. Two of these
indicators with very different properties were used in this work: the
\Halpha line and the FIR continuum emission.

\subsubsection{The H$\alpha$ line}
\label{sec:halpha-line}
Star formation rates, $\psi$, for all the galaxies in the primary catalogue
were estimated from \Halpha luminosities, $L_{\rm H\alpha}$, using the
relation presented in \cite{1998ARA&A..36..189K}:
\begin{equation}
\psi ({\rm M_{\odot}\,yr^{-1}}) = 
        \frac{L_{\rm H\alpha}}{1.27\times10^{34}\,{\rm W}}.
\end{equation} 
This relation is valid in the case of optically thin recombination,
and when the star formation obeys the \cite{1955ApJ...121..161S}
initial mass function.

\Halpha fluxes were calculated from line parameters measured by the
SDSS spectroscopic data processing pipeline, outlined in
\cite{2002AJ....123..485S}. As discussed above, while our sample is
drawn from DR1, we used the DR2 spectroscopy data-products which are
better calibrated and more robust.  The pipeline uses a number of
absorption and emission lines, and broader features, to determine
redshifts from spectra. Therefore, the \Halpha\ line need not be
prominent to be measured and, indeed, in the majority of objects in
our volume limited sample it is measured below the noise level or in
absorption.

Corrections for dust obscuration, stellar line absorption and the
3\inarcsecs\ aperture were applied following the prescription
presented in \cite{2003ApJ...599..971H}. Both the \Halpha\ and \Hbeta\
lines were corrected for stellar absorption. The equivalent width
corrections applied were 1.3\,\AA\ and 1.65\,\AA\ respectively.  As
discussed in \cite{2003ApJ...599..971H}, these values are lower than
the true stellar absorption because SDSS spectra resolve the
absorption profile.  Dust absorption correction was calculated from
the observed \Halpha/\Hbeta ratio, assuming an intrinsic ratio of 2.86
and the extinction curve presented in \cite{1989ApJ...345..245C}.  If
the \Halpha line was measured with a signal to noise ratio of less
than 2, or if it was measured in absorption, no extinction correction
was applied.  If the measured \Hbeta\ flux failed these criteria, the
extinction was calculated assuming an upper limit for the \Hbeta\ flux
of $2.3\times10^{-19}\,\unit{W}\,\unitp{m}{-2}$ which is approximately
the detection limit.

Finally, the \Halpha fluxes were corrected for the
3\inarcsecs-diameter circular aperture of the fibres, considerably
smaller than most of the galaxies in our sample. The correction was
done by scaling the \Halpha flux by the ratio of the total
\emph{r}-band flux to the \emph{r}-band flux measured within the fibre
itself. This procedure has been shown, by \cite{2003ApJ...599..971H},
to provide excellent quantitative agreement with star-formation
estimates from global galaxy properties such as radio, far-infrared
and {\it u-}band continuum emission. This procedure assumes that the
measured \Halpha\ equivalent width is representative of the whole
galaxy. For the volume-limited sample the median physical size probed
by the 3\inarcsecs\ aperture is 4.4\,kpc. The fibre therefore probes a
significant fraction of the galaxy and this accounts for the reported
accuracy in \cite{2003ApJ...599..971H}.

\subsubsection{Far-infrared continuum }

Far infrared traces recent star formation, where the natal dust clouds
still surround the young massive stars and re-processes most of their
bolometric output.  The effects of obscuration at these wavelengths
are far less prominent than at optical/UV wavelengths, removing a
potential source of systematic errors.

Flux densities at 60\,\micron\ and 100\,\micron\ were obtained for
each galaxy in our volume limited catalogue by automated query of the
SCANPI service\footnote{Available at
http://irsa.ipac.caltech.edu/applications/Scanpi/. Provided by NASA/
IPAC Infrared Science Archive, which is operated by the Jet Propulsion
Laboratory, California Institute of Technology, under contract with
the National Aeronautics and Space Administration.}, which co-adds
individual \emph{IRAS\/} scans at the position of each of the
galaxies. The measurement in each band was rejected if the reported
signal to noise was less than 2.5 or if the template correlation
factor was less than 0.9. The FIR luminosity to SFR calibration we
applied was the same as used by \cite{2003ApJ...599..971H}, namely:
\begin{equation}
\psi_{\rm FIR}\,({\rm M}_{\odot}\,{{\rm yr}^{-1}}) = \frac{f L_{\rm
    FIR}}{1.39\times10^{36}\,\unit{W}},
\end{equation}
where
\begin{equation}
  f=\left\{ 
    \begin{array}{cc}
      
      1+\sqrt{\frac{2.186\times 10^{35}\,\unit{W}}{L_{\rm FIR} }} & 
	L_{\rm FIR} > 2.186\times 10^{37} \\

	0.75\left( 1+\sqrt{\frac{2.186\times10^{35}\,\unit{W}}{L_{\rm
	FIR}}}\right) & 
	L_{\rm FIR} \leq 2.186\times10^{37} .
    \end{array}\right.
\end{equation}
The effective beam size of \emph{IRAS\/} observations is approximately
2 arcminutes, much larger than the spectroscopic aperture; for this
reason the FIR data suffer from the opposite problem to that suffered
by the \Halpha\ data -- namely, that besides accounting for the entire
FIR output of the primary galaxy, the measured FIR may be
significantly elevated by the output of other galaxies, most notably
the identified companion.

\subsubsection{Specific Star Formation Rates}
\label{sec:ssfr}

The star formation rates calculated above are not ideal for study of
the triggering of star formation as they scale with galaxy mass. A
more suitable measure is the specific star formation rate, defined as
the SFR divided by the stellar mass. Approximate stellar masses were
calculated from \zband\ ($\lambda_{\rm eff} = 9097\,{\textrm{\AA}}$)
luminosities, using $M_{z} ( \odot ) = 4.52$, as used by
\cite{2001AJ....122.1104Y}, and assuming one solar luminosity
corresponds to one solar mass. The Hubble parameter $H_{0} = 70
\,\unit{km}\,\unitp{s}{-1}\,\unitp{Mpc}{-1}$ is used throughout, for
consistency with \cite{2003ApJ...599..971H}.

\begin{figure}
\includegraphics[clip,width=\columnwidth]{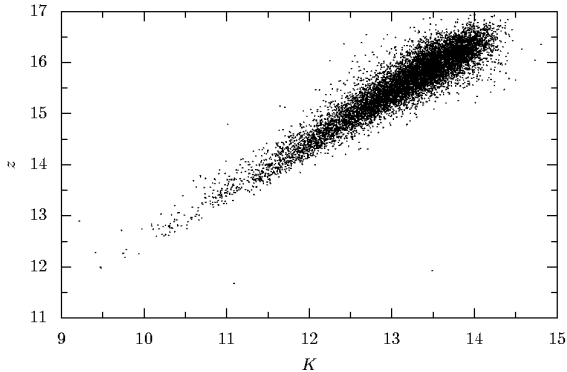}
\caption{A scatter plot of $K$- vs. $z$-band magnitudes for the
  galaxies in the volume-limited sample detected in the 2MASS XSC.}
\label{fig:kz}
\end{figure}

To calculate the \zband\ luminosity, we used the reported Petrosian
magnitudes as introduced by \cite{1976ApJ...209L...1P} and discussed
extensively in the context of SDSS by
\cite{2002AJ....124.1810S}. Study by \cite{2000A&A...361..451S},
although of limited size, suggests that the radial light profile of
interacting galaxies is not significantly extended compared to
non-interacting galaxies.  Even if there was a systematic extension,
since the fraction of galaxy light measured by Petrosian magnitudes
depends only on the shape of the light profile of the galaxy and not
on the surface brightness \citep{2002AJ....124.1810S}, if the effect
of tidal interaction on near-infrared galaxy light distribution is to
only change its characteristic length scale, there should be no bias
in estimated masses of interacting and non-interacting systems.

The \zband\ light will inevitably suffer some intrinsic extinction. To
asses whether it is an un-biased estimator of galaxy mass we have
correlated our volume-limited sample with the Two Micron All Sky
Survey (2MASS) Extended Source Catalogue (XCS) which is described in
\cite{2000AJ....119.2498J}. In Figure~\ref{fig:kz}, we compare the
Petrosian \zband\ magnitudes with total $K$-band magnitudes for the
subset of galaxies which were detected in the XSC -- there is a good
correlation between them with a dispersion around proportionality of
0.25 magnitudes. Since $K$-band magnitudes are generally believed to
give an extinction independent estimate of galaxy mass we conclude
that for this sample the \zband\ magnitude also provides an un-biased
estimate.  As a further check, we have investigated this correlation
for close pairs ($r_{p}<50\,\unit{kpc}$) and find it is essentially
unchanged from that shown in Figure~\ref{fig:kz} with a measured
dispersion of 0.27 magnitudes.

\subsection{Defining a Catalogue of Galaxy Pairs}
\label{sec:defin-catal-galaxy}

To investigate the effects of galaxy interactions, we have identified
the nearest companion galaxy to each member of our primary volume
limited catalogue. The method used for this was influenced by the
selection constraints of the spectroscopic follow up sample of the
SDSS. In particular, the 55\inarcsecs\ minimum separation between
targets on any one plate introduces a significant bias against finding
close pairs of galaxies if only galaxies with measured spectra are
considered. Therefore, we base our search for pairs on all the
galaxies identified in the imaging data, using measured redshifts
where available.

The procedure used was as follows: for each galaxy in the primary
catalogue, a list of candidate companions out to a projected
separation of 300\,kpc was assembled from the SDSS imaging galaxy
catalogues. Each of these candidates were then, in the order of
increasing separation, evaluated according to following rules until a
satisfactory match was found:
\begin{itemize}
\item If the angular separation between the primary galaxy and the
  candidate was less than three times the \rband\ Petrosian half-light
  radius of the primary (the mean of this value for the volume limited
  sample is 9\inarcsecs), the candidate was rejected. This is
  necessary to remove the small number of cases where the photometric
  pipeline incorrectly de-blends single galaxies into multiple
  components.
  
\item If the candidate has a measured redshift, the recessional
velocity difference is evaluated: if it is less than $900\,\kms$ the
candidate is accepted, otherwise it is rejected. This is a relaxed
constraint. which, according to
\cite{2000ApJ...536..153P,2002ApJ...565..208P}, includes all
potentially interacting systems and avoids a bias toward dynamically
bound systems.
\item If no redshift was measured, the \zband\ magnitude difference
  between the candidate and the primary galaxy is evaluated: if it
  less than 2 magnitudes the candidate was accepted, otherwise it was
  rejected.
\end{itemize}
This procedure resulted in 12492 of the 13973 galaxies in the volume
limited sample being identified with companions.  Of these, 2038 have
measured spectra, whilst the remaining systems have only imaging data.

\subsubsection{Validity of Spectroscopic-Photometric Pairs}

As described above, candidate pairs were not restricted to the MGS,
but were taken from all the galaxies identified in the imaging
survey. However, in the absence of redshift information, photometric
properties must be used to determine if the pairing is just a
projection effect or if the galaxies are physically close.

We adopted a relatively simple criterion, comparing the Petrosian
\zband\ magnitudes of the two galaxies. This technique suffers both
from incompleteness, when there is a large intrinsic luminosity
difference between true companions, and contamination by background
galaxies. As both of these effects depend sensitively on the cut-off
brightness ratio, we now justify the choice of this quantity.

\begin{figure}
\includegraphics[clip,width=\columnwidth]{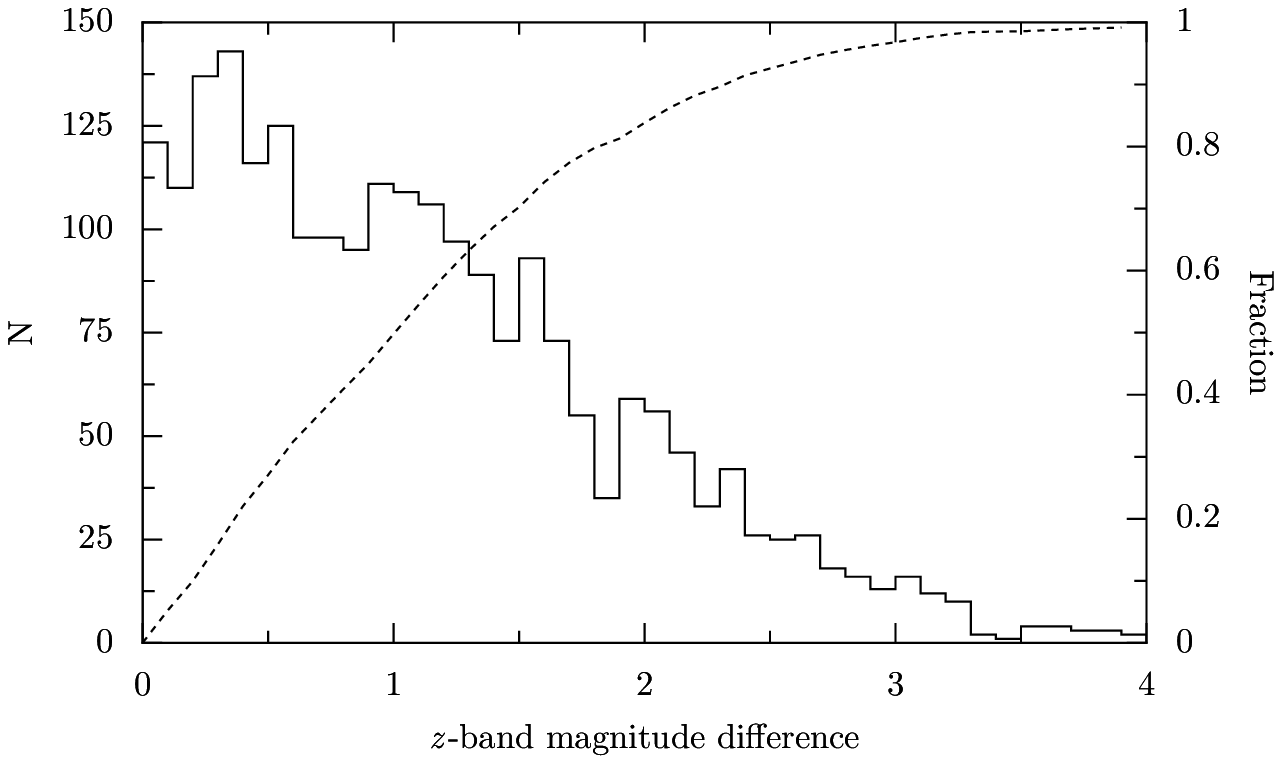}
\caption{The distribution (solid line) and cumulative distribution
  (dashed line) of \zband\ magnitude differences in spectroscopically
  confirmed pairs.}
\label{fig:zzpair-magdif}
\end{figure}

Our choice of $\Delta m_{z} = 2$ implies that companions which are
fainter than the primary galaxy by more than about a factor of 6.3 in
the \zband\ will be rejected, unless they have a measured
redshift. The \zband\ was chosen for this comparison as it most
closely traces the mass of the galaxy. To assess the significance of
this incompleteness we have calculated the distribution of \zband\
magnitude differences in all pairs which have measured spectra, and
therefore do not suffer from the above effect. The results, presented
in Figure \ref{fig:zzpair-magdif}, show that $\Delta m_{z} = 2$ is in
the tail of the distribution and that the completeness for this choice
of $\Delta m_{z}$ is approximately 80\%.

We have also investigated the probability that a detected pair is due
to background contamination as a function of the maximum magnitude
difference. To do this we have used the galaxy count from
\cite{2001AJ....122.1104Y}. The parameterised integrated counts, where
$m$ denotes the \zband\ magnitude, are

\begin{equation}
  N(\textrm{{\it z\/}-mag}<m)= \frac{11.47}{\textrm{deg}^{2}} \times 10^{0.6 (m-16)}.
\end{equation}

\begin{figure}
\includegraphics[clip,width=\columnwidth]{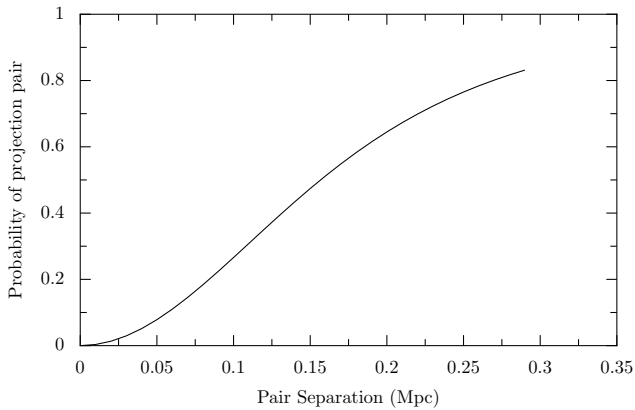}
\caption{Mean probability of a non-physical pair as a function of
  separation from the primary galaxy.}
\label{fig:project-p}
\end{figure}

The probability of a detected pair being due to contamination is then
calculated as a function of physical separation and averaged over the
whole of the primary catalogue. The results are presented in Figure
\ref{fig:project-p}. They show that at projected separation of over
150\,kpc, contamination of our pair sample by background or foreground
galaxies becomes a significant problem. Therefore, our results at
separations greater than 150\,kpc are significantly diluted by
non-physical pairs; the reported separations in this regime can,
however, be regarded as reliable lower limits of separation to the
true nearest companion.

\subsection{Concentration Index}
\label{sec:concentration-index}

Galaxy morphology has long been known to correlate strongly with the
star-formation rate
\citep[e.g.,][]{1998ARA&A..36..189K}. Additionally, numerical
simulations by \cite{1996ApJ...464..641M} show that the time
dependence of gas inflow during interactions, and the star formation
which results, depend sensitively on the structure of the progenitor
galaxies, and in particular on the bulge to disk mass ratio. In order
to measure accurately the effect of interactions on star formation, it
is important to consider these effects.

The traditional classification of galaxy morphology is the Hubble
sequence. There is a well established correlation between \Halpha
equivalent width and Hubble type, equivalent width increasing from
zero (within observational errors) in E/S0 galaxies to 20--150\,\AA\
in late-type spiral and irregular galaxies, with the same trend shown
between Sa through to Sc spirals
\citep[see][]{1983AJ.....88.1094K}. The Hubble classification is based
on three characteristics: the bulge-to-disk light ratio, tightness of
spiral arms and degree of resolution of spiral arms. Besides being
hard to automate, the Hubble classification is non-ideal for studies
of star formation as the last of the three criteria in particular
depends significantly on star formation. That is, the spiral arms are
more likely to be resolved if there is significant star formation
along them.

In this work we adopt an alternative morphological classification
scheme, the concentration index, which is described in detail by
\cite{1958PASP...70..364M}, \cite{1984ApJ...280....7O}, and
\cite{1994ApJ...432...75A}. The concentration index, $C$, is defined
as the ratio of Petrosian 50\%- to 90\%-light radii as measured in the
\rband. Low values of this quantity correspond to systems with high
central concentrations of light which in turn are of an early
morphological type. The correlation between classical morphological
classification and the concentration index for a sample of galaxies
taken from the SDSS is presented in \cite{2001AJ....122.1238S}.

We use the concentration index both to exclude early-type systems from
our study of triggered star-formation and to examine the efficiency of
this triggering as a function of morphological type. For the purpose
of removing early-type systems, we exclude all galaxies with a
concentration index less than 0.375, larger than the 0.33 value
proposed by \cite{2001AJ....122.1238S}.  In doing so, we reduce
contamination of our late-type sample to less than 5\% \citep[see
Figure 11 of ][]{2001AJ....122.1238S}, at the expense of completeness.

\subsection{Pair Morphology}

In the later stages of interaction it is often not possible to resolve
the nuclei of interacting galaxies, especially in visible light, and
even when they are resolved the automated algorithm in the SDSS
photometric pipeline might not de-blend them. More importantly, the
algorithm used to select nearest companions
(Section~\ref{sec:defin-catal-galaxy}) rejects all objects closer than
three times the Petrosian half-light radius. As a result, our pair
sample does not contain the closest pairs or merging objects.  At the
median redshift, and using the median half-light radius, the minimum
galaxy separation corresponds to 14\,kpc.

To assess the importance of this effect, we have visually inspected
three sub-samples, each containing approximately 30 galaxies. The
first consisted of the most actively star-forming objects, the second
of medium star-forming objects and the third was a random control
sample. In each case we evaluated the possibility that the galaxy is
in the process of gravitational interaction via the presence of tidal
arms and double nuclei. The results are presented in
Section~\ref{sec:res-pair-morphology}.

\section{Results}

\begin{figure}
   \includegraphics[clip,width=\columnwidth]{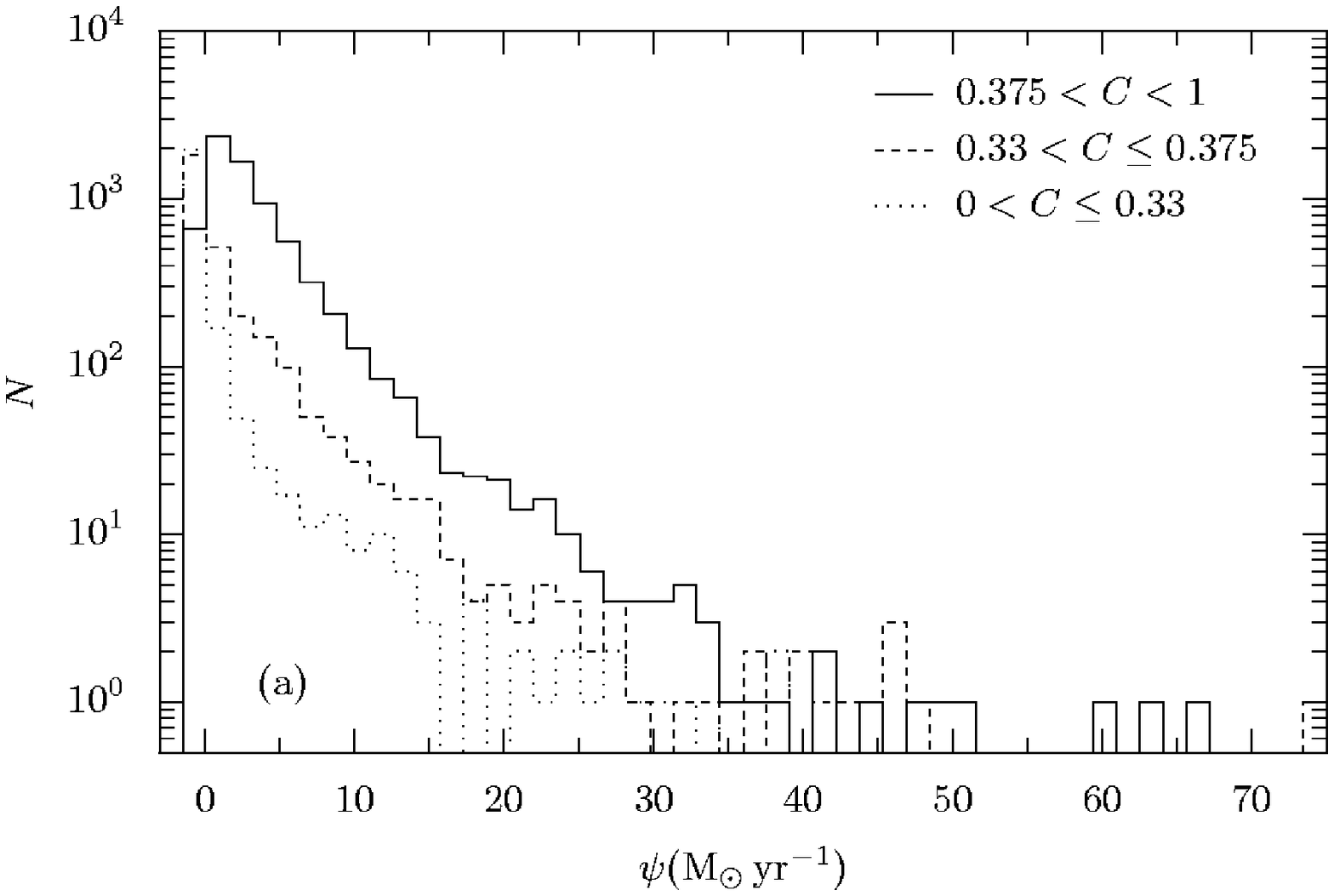}
   \includegraphics[clip,width=\columnwidth]{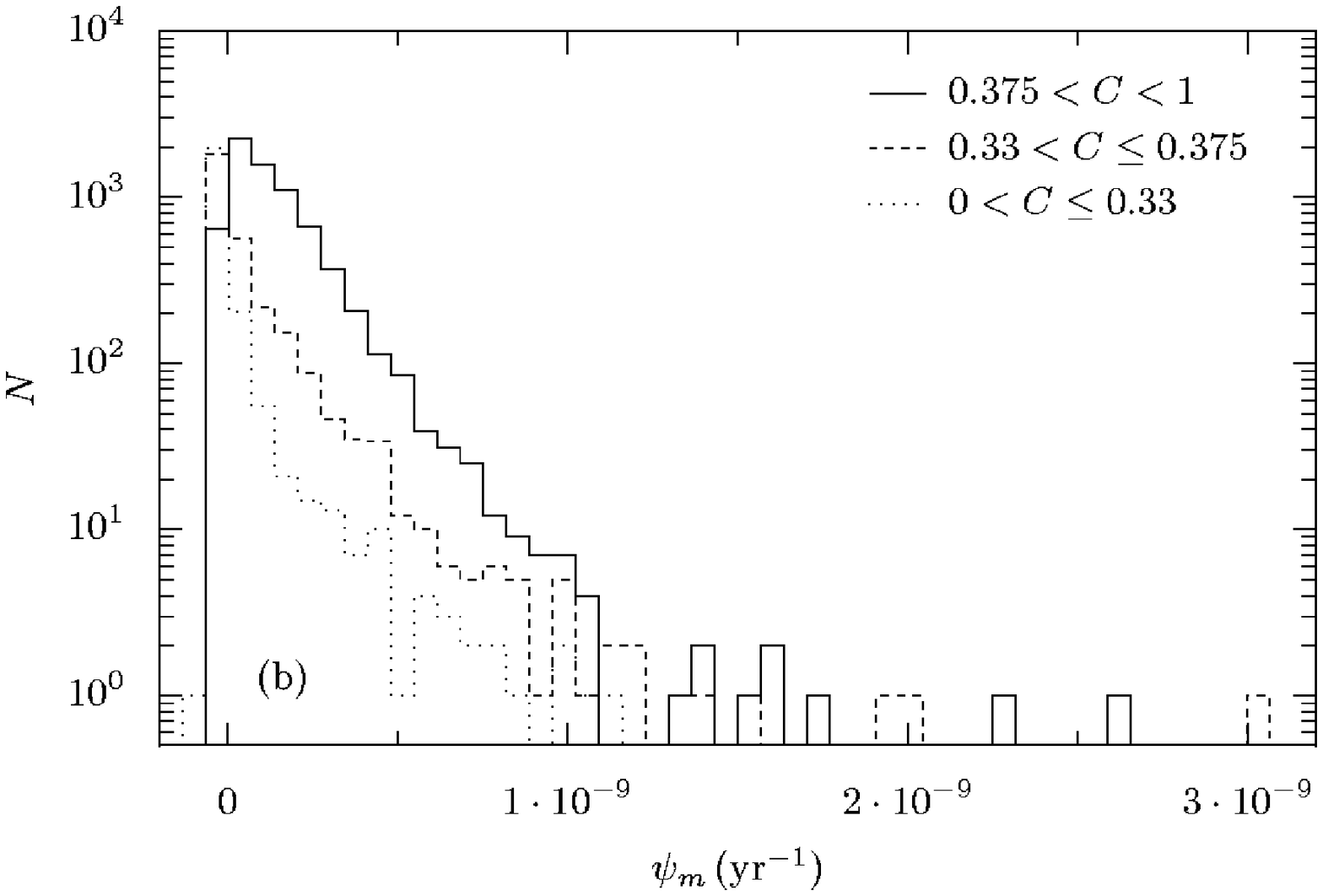}
   \caption{The distribution of (a) absolute and (b) specific
     star-formation rates for galaxies in the volume limited sample,
     split according the concentration index into three subsamples.}
   \label{fig:sfr-dist}
\end{figure}

Our primary sample consists of 13973 galaxies of which 12492 have an
identified companion.  The distribution of morphological types, as
determined by the concentration index
(Section~\ref{sec:concentration-index}), of the the galaxies in the
primary sample, together with their companion, is shown in
Table~\ref{tab:stats_table}.  The distribution of absolute star
formation rates, $\psi$, and specific star formation rate, $\psi_{m}$,
are shown in Figure~\ref{fig:sfr-dist}.  Here we distinguish between
three morphologically-defined sub-samples: late-type ($C>0.375$),
early-type ($C \leq 0.33$) and mixed ($0.33<C\leq0.375$). As expected,
given the good correlation between concentration index classification
and Hubble type, the vast majority of star-forming systems are
classified as late-type.  The mean star formation rates are 0.7, 1.7,
and 3.3\,\MSolar\,\unitp{yr}{-1} for the early, mixed and late type
sub-samples respectively.

In this and following sections, we consider each pair to consist of a
primary galaxy and companion and examine the star-formation rate of
the primary galaxy in terms of the properties of the interacting pair.

\ctable[
  cap={Distribution of morphological types},
  caption={Distribution across concentration index classes of galaxies
  in the volume limited sample and their companions.},
  star,
  label=tab:stats_table]
       {cccccc}
       {\tnote[a]{The number of galaxies with no companion identified
	   within the  300\,kpc  maximum search radius.}}
       {\FL
	 Primary Galaxy	& Total & \multicolumn{3}{c}{Companion Galaxy} & No
	 companion\tmark[a] \NN
	 \multicolumn{2}{c}{}		& $C> 0.375$	& $0.33 <C \leq 0.375$ & $C\leq 0.33$ \ML
	 $C> 0.375 $		& 7936	& 5120	& 1500	& 555 & 761  \NN
	 $0.33 <C \leq 0.375$	& 3344	& 2064 	& 681	& 270 & 329  \NN			
	 $C\leq 0.33$		& 2693  & 1455  & 576	& 271 & 391  \ML
	 Total			& 13973 & 8639  & 2757  & 1096& 1481 \LL
       }

\subsection{Star formation as function of projected separation}

\begin{figure}
\includegraphics[clip,width=\columnwidth]{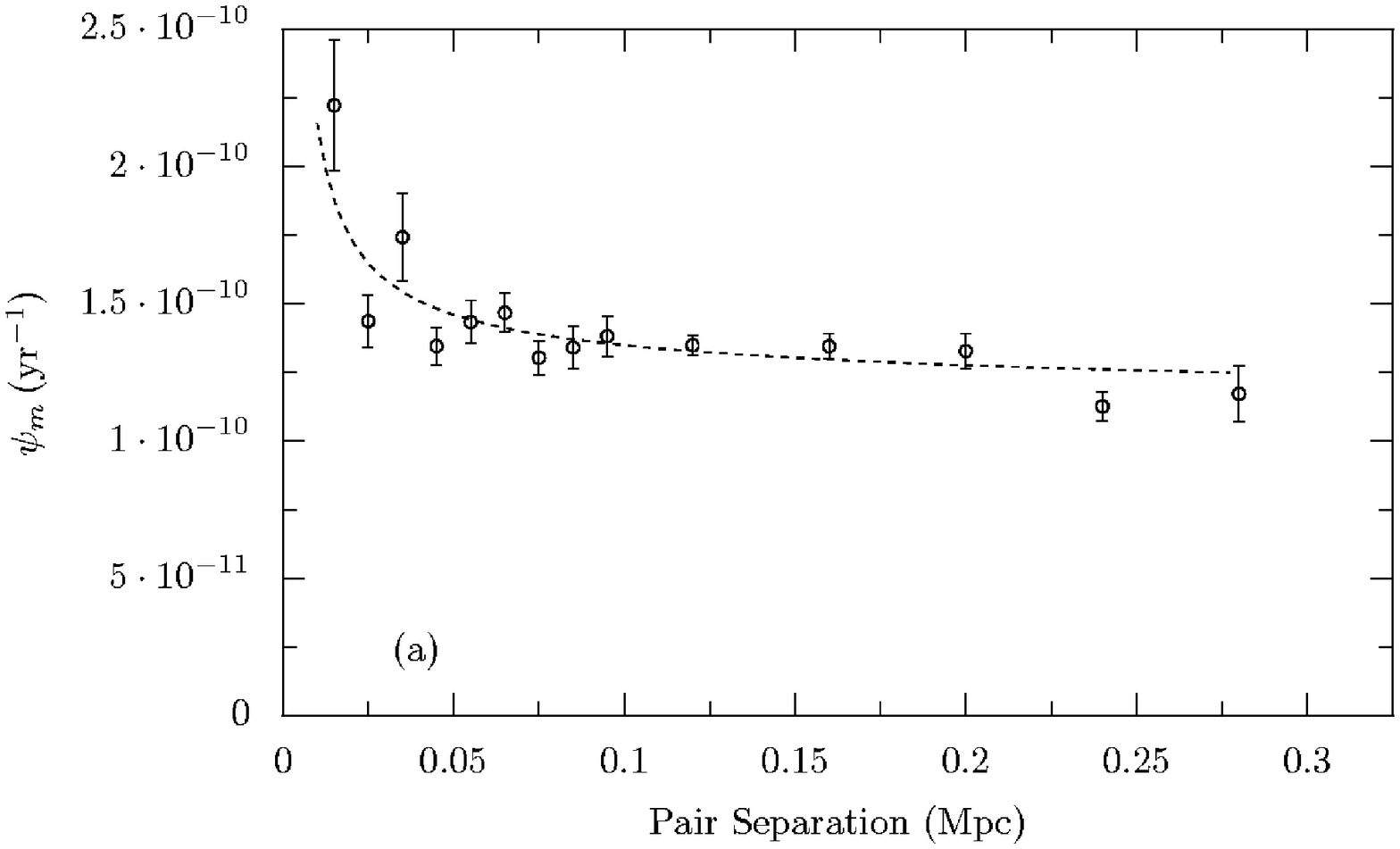}
\includegraphics[clip,width=\columnwidth]{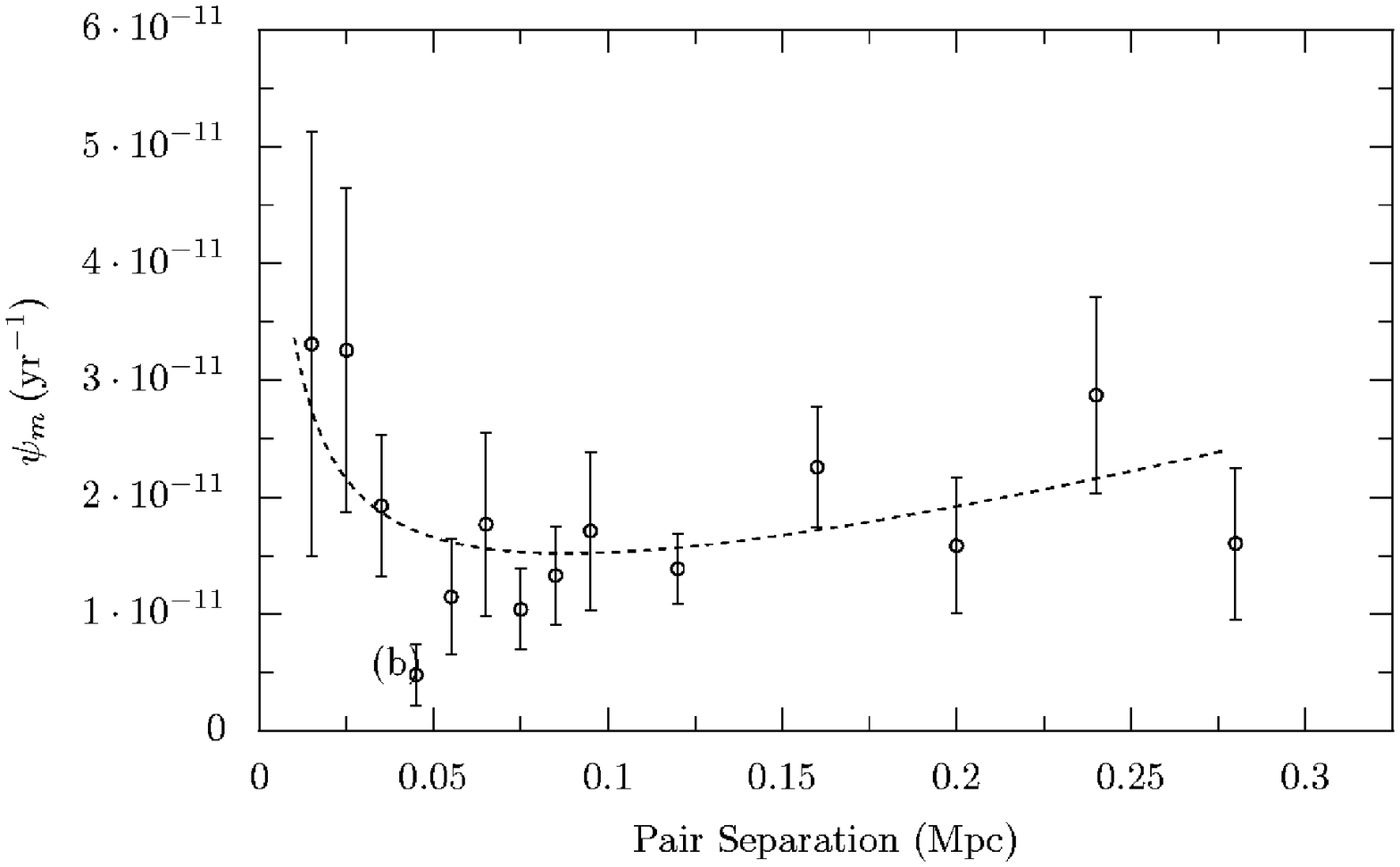}
\includegraphics[clip,width=\columnwidth]{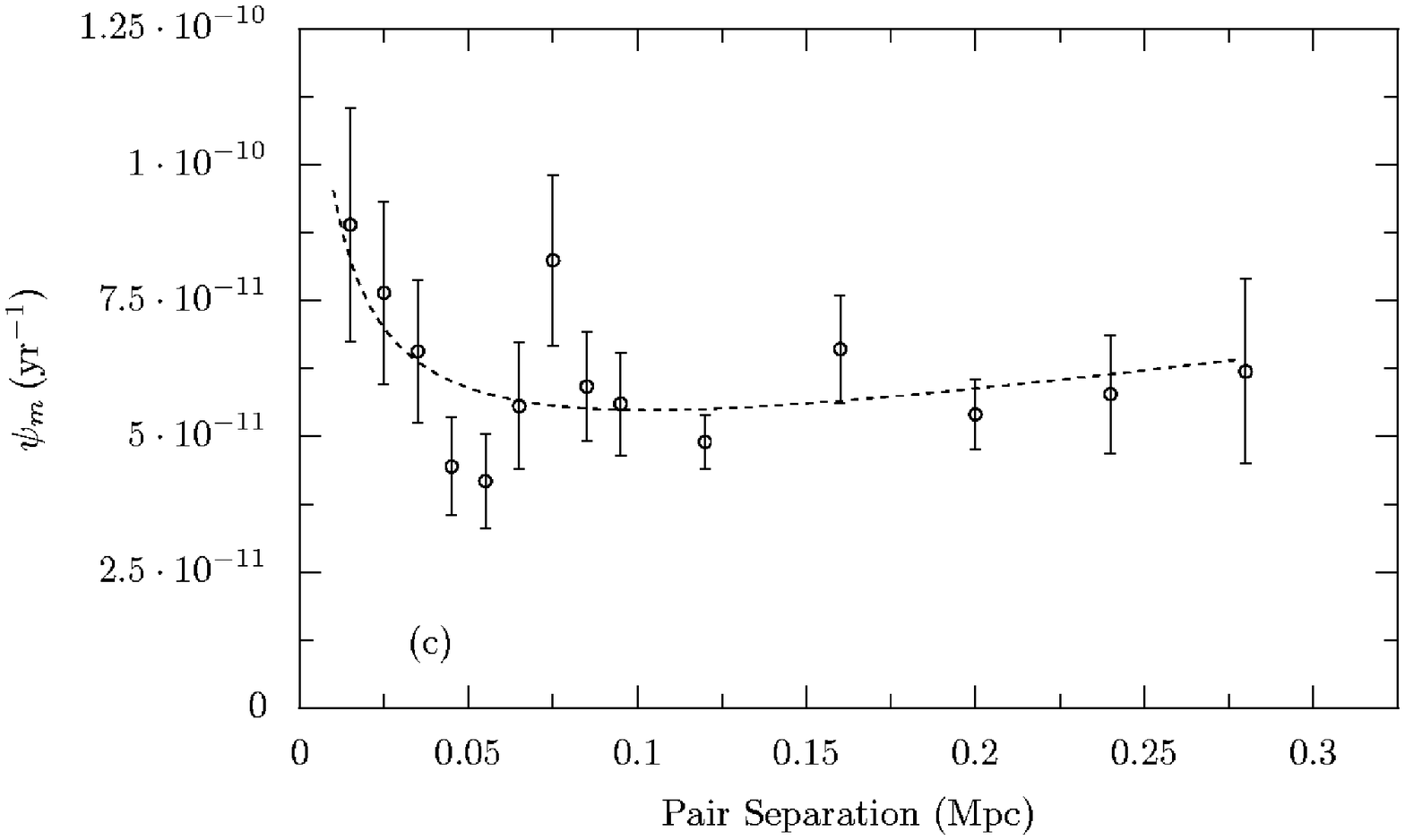}

\caption{Specific star formation rates as a function of pair
  separation for late (a), early (b), and mixed (c) galaxy types.}
\label{fig:psim_DD}
\end{figure}
\ctable[
  cap={Power law dependence of mean SSFR on projected separation for
 late-, early- and mixed-types}, caption={The dependence of the
 specific star formation rate on projected separation is characterised
 by fitting a two power-law function to the specific star
 formation rate vs pair separation data (Fig. \ref{fig:psim_DD}) after
  its been divided according to morphological type.},star,
  label=tab:fits_Cbins]
       {ccccc}
	{}
       {\FL
	 Concentration & $a_{1}$ 	&
  $\beta_{1}$	& $a_{2}$  		& $\beta_{2}$  \NN
Index bin & $(10^{-11}\,\unitp{yr}{-1})$ & & $(10^{-11}\,\unitp{yr}{-1})$ & \ML
	 $0<C\leq0.33$          & $2.04$	& $-0.53$ & $0.11$ 	& $1.19$ \NN
	 $0.33<C \leq 0.375$	& $6.30$	& $-0.41$ & $0.70$	& $0.73$ \NN		  
         $0.375<C \leq1.0$ 	& $13.6$        & $-0.04$ & $2.90$&  $-1.04$   \LL
       }

To investigate the dependence of specific star formation rate,
$\psi_m$, on true pair separation we consider firstly the dependence
of $\psi_m$ on projected separation, looking subsequently at the
dependence on velocity separation.  In Figure~\ref{fig:psim_DD}, we
plot the mean specific star-formation rate against projected pair
separation.  The primary galaxy in each pair has been assigned, using
the concentration index, to one of three morphological classes
representing late-, mixed- and early-types (see
Table~\ref{tab:stats_table}).  The data have been binned into bins
10\,kpc wide for separations less than 100\,kpc and bins 40\,kpc wide
for separations greater than 100\,kpc.  For each bin we have indicated
the mean and the estimated statistical variance of the mean; the
variance is dominated by sample variance rather than measurement
errors.  In each case we characterise the variation of $\psi_{m}$ with
pair separation, $r_{p}$, by fitting a simple two power-law function
to the un-binned data:
\begin{equation}
\psi_{m}(r_{p}) = a_{1} \left(
\frac{r_{p}}{25\,\unit{kpc}}\right)^{\beta_{1}} + 
a_{2} \left( \frac{r_{p}}{25\,\unit{kpc}}\right)^{\beta_{2}}.
\end{equation}
The best-fitting parameters are detailed in Table~\ref{tab:fits_Cbins}.

For each morphological class, there is a significant increase in the
specific star-formation rate at projected separations less than
30\,kpc.  For late-type systems there is some indication that the mean
specific star-formation rate decreases systematically with projected
separation out to separations of about 300\,kpc where it reaches a
value of $1.2\times 10^{-10}\,\unitp{yr}{-1}$.  Those systems
classified as early- and mixed-types only show significant enhancement
for projected separations less than about 25\,kpc.  The greater
dispersion in these plots reflects the larger intrinsic range in the
estimated star-formation rates in these systems, many of which are
effectively zero.

\begin{figure}
\includegraphics[clip,width=\columnwidth]{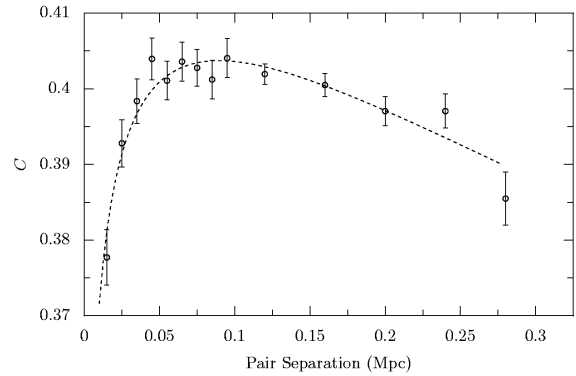}
\caption{Mean concentration index, $C$, as a function of pair
  separation for the full volume limited sample. The dashed line is
  the least-squares best-fitting function of the form $C=a_{1} (
  r_{p}/25\,\unit{kpc})^{\beta_{1}} + a_{2}(
  {r_{p}}/25\,\unit{kpc})^{\beta_{2}} $; the best fit parameters are:
  $a_{1}=1.6$, $\beta_{1}={0.19}$, $a_{2}=-1.20$, and
  $\beta_{2}=0.24$.}
\label{fig:comp-sep}
\end{figure}

The use of the concentration index as a morphological classifier does
introduce some potential problems.  Any enhancement in nuclear surface
brightness of a galaxy will reduce the value of the concentration
index for that system.  In particular, an intense nuclear starburst
may result in a late-type system being placed in the wrong
morphological class.  To examine this effect further we show in
Figure~\ref{fig:comp-sep} the mean concentration index for all systems
as a function of projected separation.  The concentration index peaks
at a separation of around 75\,kpc and rapidly declines at separations
smaller than 50\,kpc.  Inspection by eye of the low concentration
index systems at small separations, strongly suggest they are disk
systems with pronounced nuclear starbursts.  The enhancement in the
specific star-formation rate at small physical separations for those
systems which we have been classified as early- and mixed-type is
likely, at least in part, to be explained by the misclassification of
disc-galaxies with strong nuclear starbursts.  We cannot eliminate the
possibility that there is a real enhancement in star formation for
intrinsic early-type systems for small physical separations to a
companion; further study of this point requires an independent means
of morphological classification.

Tidal distortion may also be contributing to the lowering of
concentration index observed in Figure~\ref{fig:comp-sep} for close
pairs. A close encounter between two galaxies can lead to stripping of
the more peripheral stellar material whilst leaving the central
stellar distribution relatively undisturbed, although the study by
\cite{2000A&A...361..451S} find no evidence for this. The effect of
such a distortion would be to increase the value of the Petrosian 90\%
light radius whilst leaving the Petrosian 50\% light radius unchanged,
reducing the concentration index. However, examination of the relative
value of the 50 and 90\% light radii with pair separation reveal both
to decline for close pair separations, indicating that nuclear
starbursts are likely to be the dominant factor in the reduction of
concentration index at close pair separations.

\begin{figure*}
\includegraphics[clip,width=\linewidth]{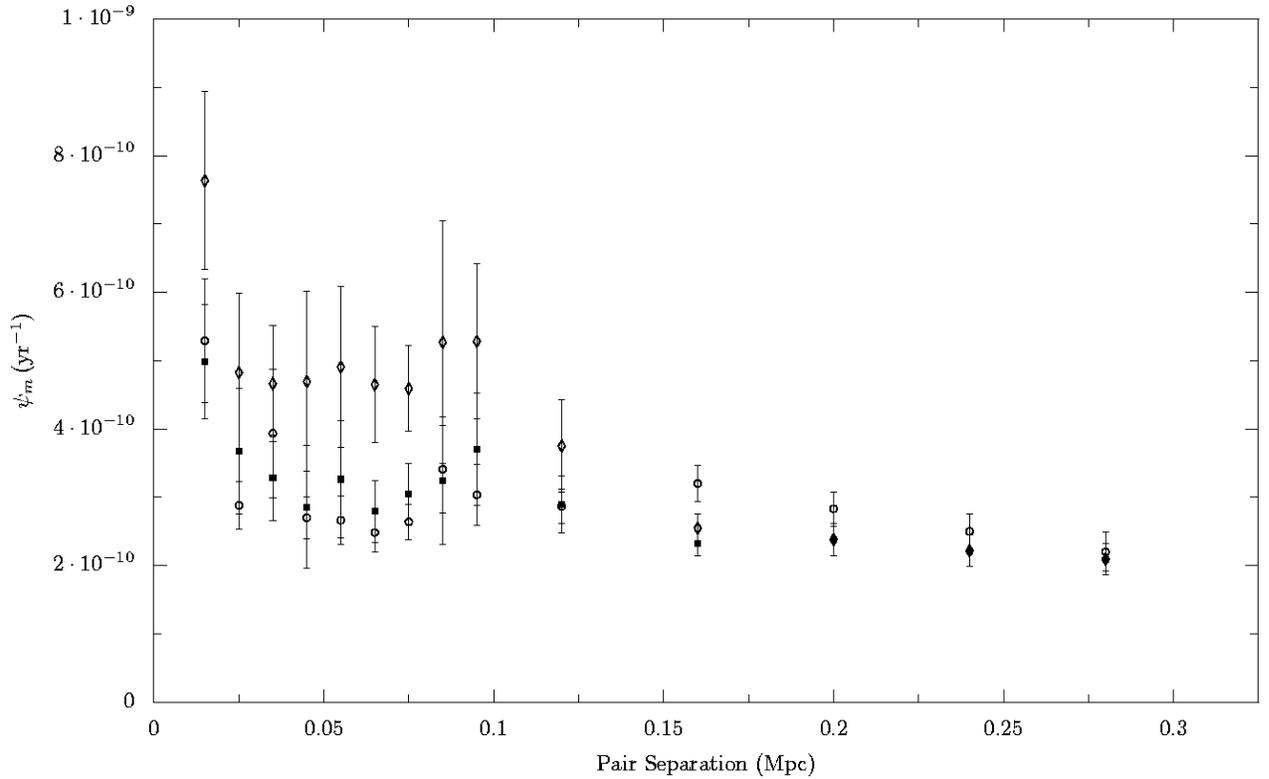}
\caption{Specific star-formation rates for galaxies detected by {\it
IRAS;} open circles: \Halpha star-formation estimator (as before);
solid squares: FIR star-formation estimator; open diamonds: FIR
star-formation estimator but normalised by mass of both the primary
and companion galaxies if they are within 1.5 arcminutes of each
other.}
\label{fig:iras}
\end{figure*}

The use of \Halpha\ luminosity as the star-formation rate estimator
involves corrections both for extinction and the finite size of the
aperture as discussed in Section~\ref{sec:halpha-line}.  These
procedures may introduce subtle biases if, for example, the effect of
tidal interaction is always to trigger a nuclear starburst.  To
investigate any such effects we compare results using the FIR and
\Halpha star-formation rate indicators -- Figure~\ref{fig:iras}.  In
order to take into account the possibility that both the primary
galaxy and its companion are contributing to the measured fluxes, two
mass normalisations of the FIR-derived star-formation rates are shown
in the figure: normalisation by mass of the primary galaxy (as
everywhere else); and, normalisation by the sum of the masses of
primary and companion galaxies if their separation is less than 1.5
arcminutes, and just the primary mass otherwise (open diamonds on the
plot).

The comparison is limited to those systems for which both FIR and
\Halpha are available so this is inevitably biased towards systems
with higher star formation rates.  However, these are precisely the
systems in which any problems associated with the use of \Halpha as a
star-formation rate estimator are likely to be the greatest.  We find
good agreement between the FIR, especially when normalised to the sum
of the masses of the primary and companion galaxies, and \Halpha data
both in terms of the trend with projected pair separation and the
magnitude of the specific star formation rate.  We conclude that the
procedure used here to estimate the star-formation rate from \Halpha
data gives a good and predominantly unbiased estimate of the true
star-formation rate.

\begin{figure}
\includegraphics[clip,width=\columnwidth]{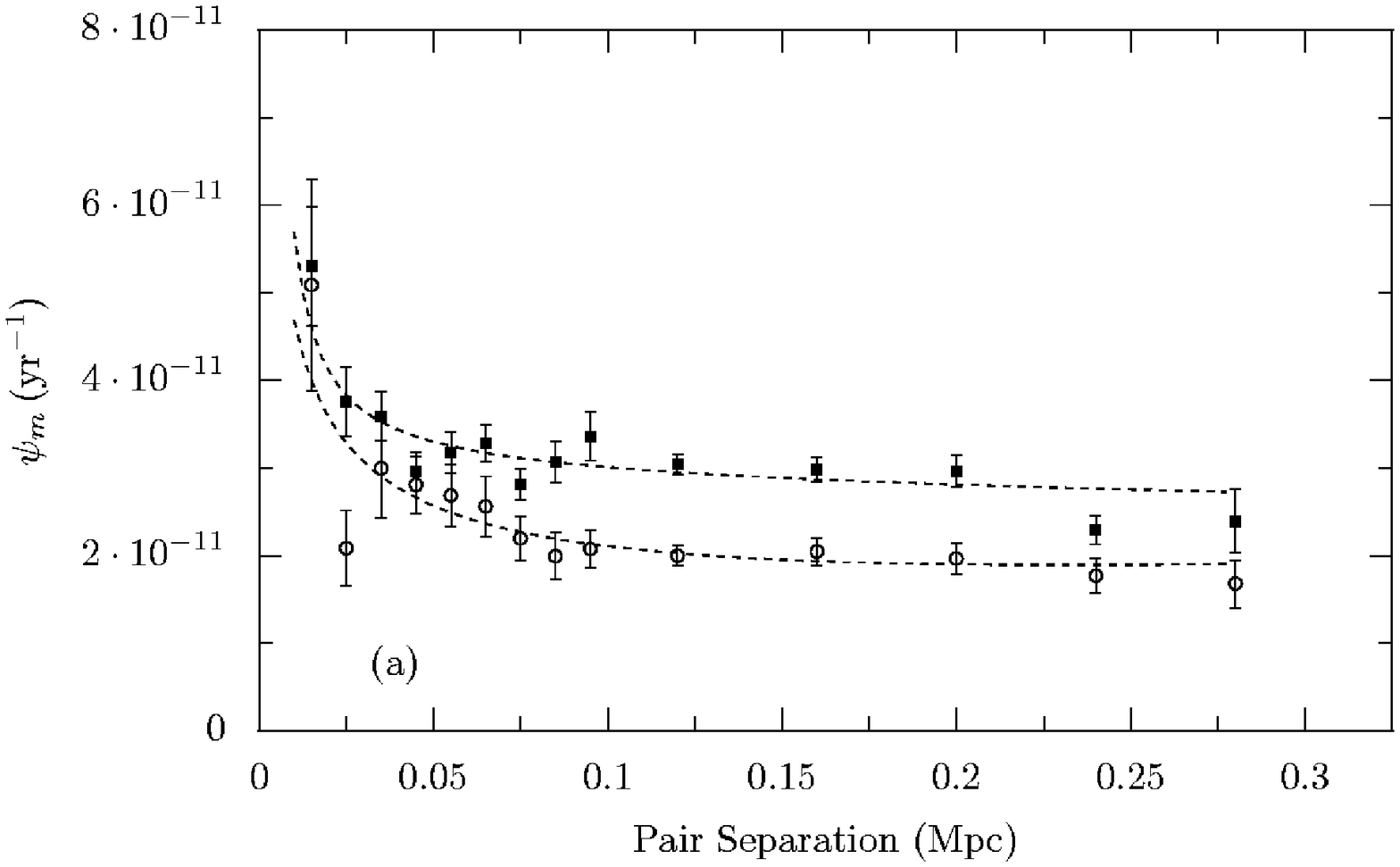}
\includegraphics[clip,width=\columnwidth]{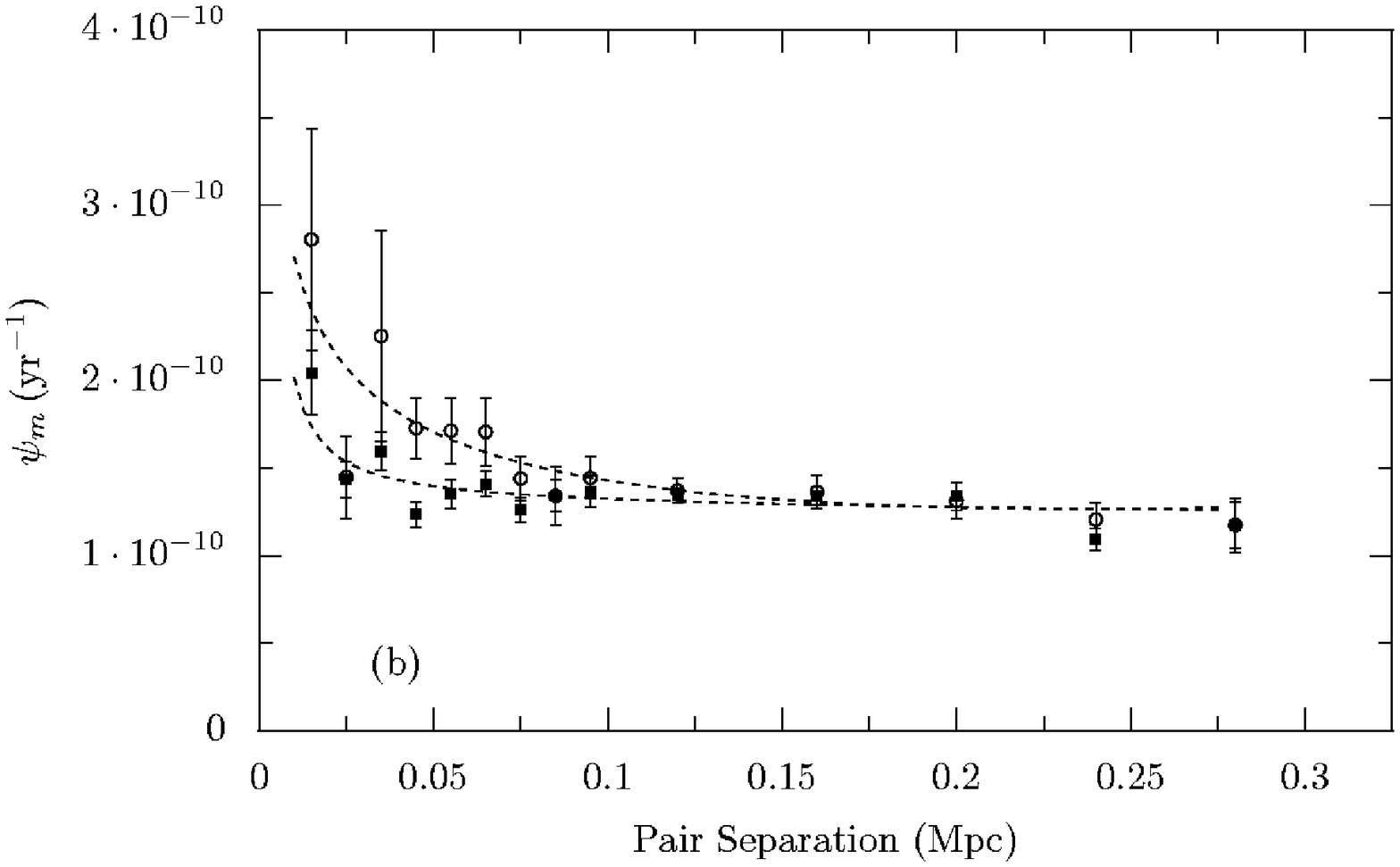}
\caption{Comparison of (a) non-aperture corrected  and (b)
  aperture-corrected star-formation rates versus separation for
  late-type galaxies in two redshift bins: $0.03<z\leq 0.07$ (open
  circles) and $0.07<z\leq0.1$ (full squares). }
\label{fig:apcoreffect}
\end{figure}

In order to further investigate the effect aperture correction has on
our results, in Figure~\ref{fig:apcoreffect} we compare non-aperture
corrected and aperture corrected star-formation rates as a function of
separation; the significance of the aperture correction is made
clearer by splitting the sample into two redshift bins. It is quite
clear that the non-aperture corrected data exhibit a similar trend to
the corrected data, the main difference being a somewhat steeper rise
of the SSFR at close separations. This is consistent with the majority
of the triggered star formation occurring in the nuclear regions of
the galaxies. Aperture correction brings the two sub-samples into good
agreement although the higher corrected SSFR of the low-redshift
subsample in the closest bin suggests that the correction may be
over-estimated in systems with close companions. The implication of
this is that galaxies with close companions have star formation is
enhanced in the nuclear region compared to the periphery -- that is,
they harbour nuclear starbursts.  Therefore although the over-estimate
of aperture correction for close pairs contributes to the uncertainty
in our measurement of enhancement of star formation in those pairs,
the \emph{existence\/} of this over-correction is in itself strong
evidence of nuclear starbursts in galaxies with close companions.

\begin{figure}
\includegraphics[clip,width=\columnwidth]{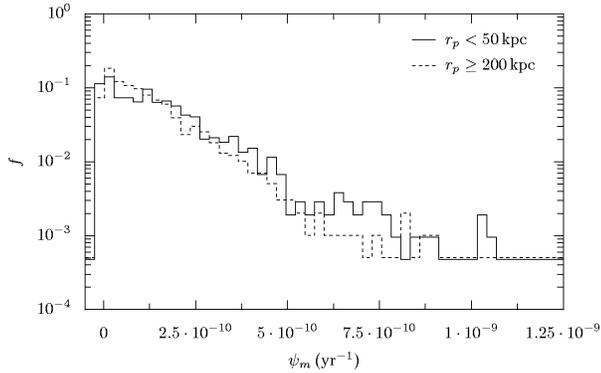}
\caption{The relative distribution of specific star-formation rates
  for galaxies from the late-type volume limited sample with close
  companions (solid line) and no companions or companions which are
  separated by more than 200\,kpc (dashed line).}
\label{fig:closepairssfrdist}
\end{figure}

\begin{figure}
\includegraphics[clip,width=\columnwidth]{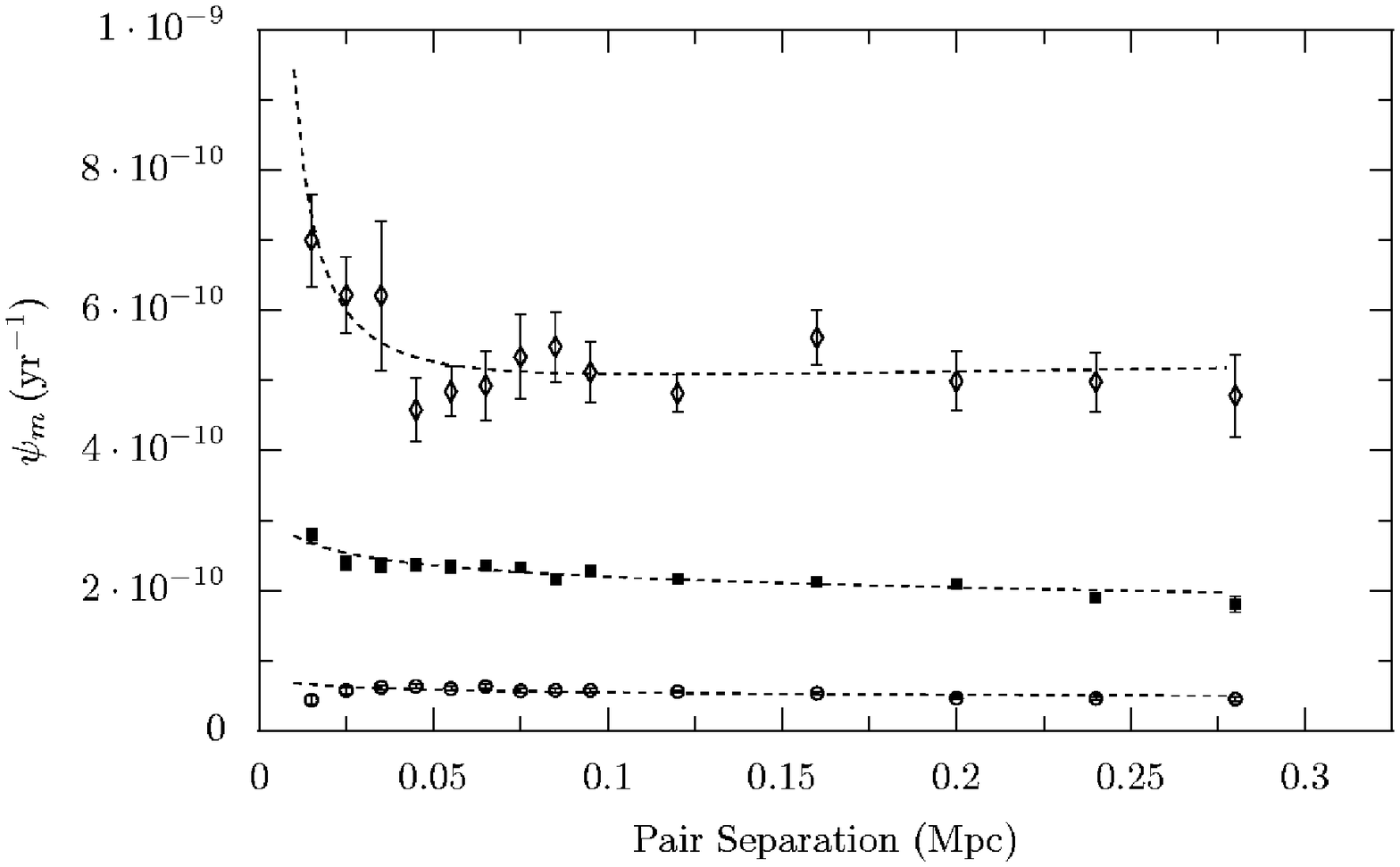}
\caption{Specific star formation versus separation for three
  sub-samples selected according to absolute star formation rate:
  $0<\psi<3\,\MSolar\,\unitp{yr}{-1}$ ,
  $3\,\MSolar\,\unitp{yr}{-1}<\psi<10\,\MSolar\,\unitp{yr}{-1}$ and
  $\psi>10\,\MSolar\,\unitp{yr}{-1}$. }
\label{fig:SFRbins}
\end{figure}

In Figure~\ref{fig:closepairssfrdist}, we show the distribution of
specific star-formation rates for two sub-samples of late type
galaxies: one of galaxies with companion within 50\,kpc projected
separation, and the other with companions at projected separation
greater than 200\,kpc or no companion at all. The Kolmogorov-Smirnov
(K-S) test indicates, at a significance level greater than 99.9\%,
that the star-formation rates in these two sub-samples are drawn from
different parent distributions.  The distributions only differ
significantly for specific star-formation rates above approximately
$3\times 10^{-10}\,\unitp{yr}{-1}$ suggesting that the observed
increase in the mean specific star-formation rate for systems with
close companions is dominated by systems showing significant increases
in their rate of star formation.  To investigate this further, we show
in Figure~\ref{fig:SFRbins} and Table~\ref{tab:sfrbin_fits} the
variation in $\psi_m$ with projected separation for late-type systems
binned according to the absolute star-formation rate: low star-forming
($0<\psi<3\,\MSolar\,\unitp{yr}{-1}$), medium star-forming
($3\,\MSolar\,\unitp{yr}{-1}<\psi<10\,\MSolar\,\unitp{yr}{-1}$), and
highly star-forming ($10\,\MSolar\,\unitp{yr}{-1}<\psi$).  The
late-type systems in the lowest star formation rate bin
$0<\psi<3\,\MSolar\,\unitp{yr}{-1}$ show no enhancement of specific
star-formation rate with decreasing pair separation.  In contrast both
of the higher star-formation rate bins
($3\,\MSolar\,\unitp{yr}{-1}<\psi<10\,\MSolar\,\unitp{yr}{-1}$ and
$\psi>10\,\MSolar\,\unitp{yr}{-1}$) show a clear increase in specific
star-formation rate with decreasing projected separation.  As is clear
from the plots, and the best-fitting parameters given in
Table~\ref{tab:sfrbin_fits}, this enhancement is most significant at
the highest star-formation rates.

\ctable[
  cap={Power law fits to SFR plots},
  caption={Parameters obtained when fitting a two power-law function
  to plots of specific star formation rate as a function of pair
  separation for three sub-samples of systems, selected according to
  SFR. },star,
  label=tab:sfrbin_fits]
       {ccccc}
	{}
       {\FL
	 SFR range\,(\MSolar\,\unitp{yr}{-1})  & $a_{1}\,(\unitp{yr}{-1})$	& $\beta_{1}$	& $a_{2}\,(\unitp{yr}{-1})$  		& $\beta_{2}$  \ML
	 $0<\psi\leq3$ & $3.26\times10^{-11}$	& $-0.10$ &	$2.98\times10^{-11}$ 	& $-0.09$ \NN
	 $3<\psi\leq10$ 	& $1.40 \times10^{-10}$	&   $-0.10$ & $1.14\times10^{-10}$	& $-0.10$ \NN		  
         $10<\psi\leq1000$ & $1.43\times10^{-10}$ & $-1.37$ & $4.03\times10^{-10}$ & $0.05$ \LL
       }

In summary, we find strong evidence for an enhancement in the specific
star-formation rate in late-type systems as a function of projected
separation to the nearest companion.  Within a projected separation of
50\,kpc there is a very significant increase in $\psi_m$ and this can
be attributed to the contribution from systems showing the largest
star-formation rates.

\subsection{Star formation properties as a function of recessional
  velocity separation}

\begin{figure}
\includegraphics[clip,width=\columnwidth]{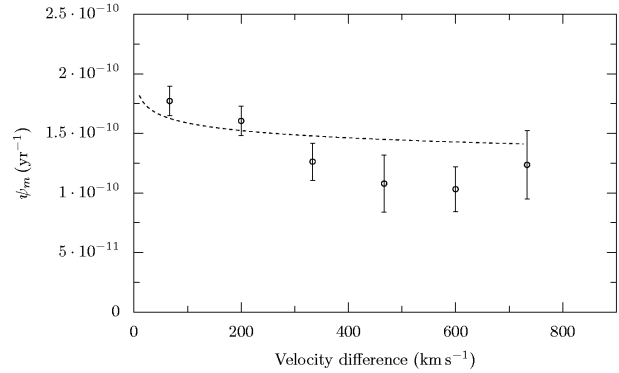}
\caption{Specific star-formation rate of late type galaxies
  ($C>0.375$) as a function of velocity separation. }
\label{fig:deltav}
\end{figure}

In approximately one-fifth of our volume limited sample the companion
galaxy has a measured SDSS spectrum, and therefore, in most cases, a
measured redshift. In Figure~\ref{fig:deltav} we examine the effect of
the recessional velocity difference on star formation rate for these
galaxies. The best-fitting parameters from fitting a
two power-law model are shown in Table~\ref{tab:deltav_fit}. The data
reveal a decline in the specific star-formation rate with increasing
recessional velocity difference out to 900\,$\kms$. The magnitude of
this effect is smaller than that observed between specific
star-formation rate and projected separation.  The steepest decline is
seen at small velocity differences.

A correlation between projected separation and velocity difference may
be expected if, for example, all systems were bound.  We find no
correlation between these two quantities for this sample; the mean
velocity separation is $\sim 200\,\kms$ and is independent of
projected separation.

\ctable[
  cap={Power law fits for deltav sample},
  caption={Parameters obtained when fitting a two power-law function
  to the plot of specific star formation rate as a function of the
  recessional velocity of spectroscopic galaxy pairs.},
  label=tab:deltav_fit]
       {cccc}
	{}
       {\FL
	  $a_{1}\,(\unitp{yr}{-1})$	& $\beta_{1}$	& $a_{2}\,(\unitp{yr}{-1})$  		& $\beta_{2}$  \ML
	  $4.56\times10^{-10}$	& $0.05$ & $1.43\times10^{-10}$ 	& $-1.38$\NN
       }

\subsection{Optical Morphology}
\label{sec:res-pair-morphology}

Optical and near-infrared imaging of {\it IRAS} selected galaxies has
revealed that the fraction of objects which are interacting/merging
increases systematically with increasing infrared luminosity.  For
example, \cite{1996ARA&A..34..749S} find this fraction increases from
about 10\% at $\log(L_{\rm FIR}/L_{\odot})$ = 10.5--11 to essentially
100\% at $\log (L_{\rm FIR}/L_{\odot}) > 12$.

As discussed in Section~\ref{sec:defin-catal-galaxy}, our procedure
for finding companions misses very close companions and actively
merging systems.  The results of the previous two sections are subject
to this bias.  In particular, systems which are undergoing strong
enhancements in their star-formation rate due to ongoing mergers are
likely to appear in our catalogue over the full range of projected
separations (and/or recessional velocity difference).  Such systems
will therefore only contribute to masking any real effect due to tidal
interactions.

We have therefore examined by eye the morphologies of 81 galaxies in
our sample.  The 81 galaxies examined were selected in three ways:
\begin{itemize}
  \item[A:] The twenty-four most highly star forming galaxies; equivalent
  infrared luminosities $\log(L_{\rm FIR}/L_{\odot}) > 11.42$. 
  \item[B:] Twenty-five
  galaxies with moderately active star formation; equivalent infrared
  luminosities in the range $11.10 < \log(L_{\rm FIR}/L_{\odot}) <
  11.13$. 
  \item[C:] A random control sample of thirty-two galaxies. 
\end{itemize}

Among the 24 galaxies in group A, 38\% show evidence of being merging
systems.  This fraction drops to 20\% in group B and 6\% in the
control sample.  The trend observed in the {\it IRAS}-selected samples
of increasing optical disturbance with increasing infrared luminosity
is therefore reflected in our volume-limited sample.  The large
fraction of merger systems at the high star-formation rates can only
reduce the observed anti-correlation between star formation and
projected separation. We discuss these effects further in
Section~\ref{sec:discussion}.

\subsection{Dependence on the properties of the interaction}
\label{sec:res-props}

We do not expect all tidal interactions to lead to enhanced star
formation.  How effective a given interaction is at triggering gas
inflow etc. may depend not only on the properties of the perturbed
galaxy (which we have examined in the previous section), but also upon
the nature of the perturber.  The nature of the perturbing galaxy can
be probed in two ways: its morphological class and mass.

\begin{figure}
\includegraphics[clip,width=\columnwidth]{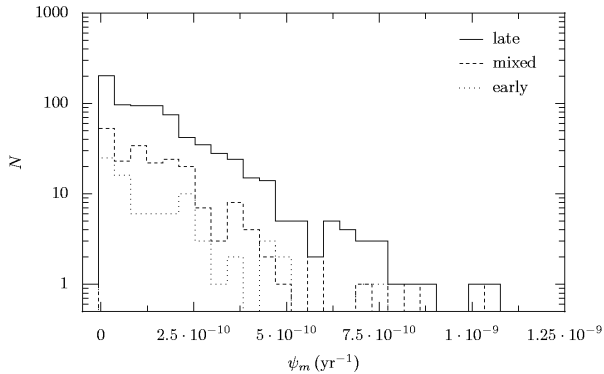}
\caption{Distribution of specific star nformation rates for galaxies
  with companions closer than 50\,kpc, split according to the
  morphological type of the companion.}
\label{fig:histpairmorph}
\end{figure}

In Figure~\ref{fig:histpairmorph} we show the distribution of specific
star-formation rates for late-type galaxies with a companion within
50\,kpc.  We split the pairs into three sub-samples depending on the
morphological type of the perturber: late-late, late-mixed and
late-early pairs.  There are no significant differences between these
sub-samples: the Kolmogorov-Smirnov (K-S) test applied between the
late-late and late-early distributions gives a formal significance of
only 20\% that they are drawn from different distributions.

\begin{figure}
\includegraphics[clip,width=\columnwidth]{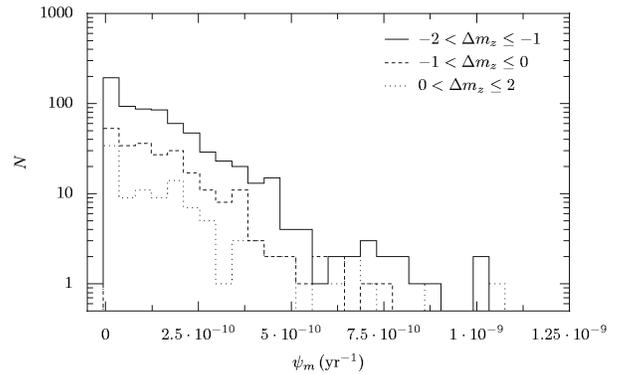}
\caption{Distribution of specific star formation rates for galaxies
with pairs closer than 50\,kpc, split according to \zband\ magnitude
difference.}
\label{fig:histmassdiff}
\end{figure}

We investigate the effect of companion mass in
Figure~\ref{fig:histmassdiff}.  Again, we split those systems which
have companions with projected separations less than 50\,kpc into
three sub-samples according to the $z$-band magnitude difference
between the primary and companion galaxy [$\Delta m_z = m_z({\rm
primary}) - m_z({\rm companion})$].  For all specific star-formation
rates we find an excess of systems where the companion is of lower
mass than the primary galaxy.  This effect is mainly due to form of
the luminosity function which rises steeply to lower luminosities at
the limit of our primary sample. Therefore, for real physical pairs, a
galaxy of a given luminosity is more likely to be interacting with a
galaxy of lower luminosity.  We have used the K-S test to compare the
SSFR of the three subsamples.  We tested the null hypothesis that the
three distributions are all drawn from the same parent distribution.
We find no evidence to reject the null hypothesis -- the formal
confidence level that the distributions are drawn from different
populations is 50\%.

\section{Discussion}
\label{sec:discussion}

The high-quality data from the SDSS has enabled us to construct a
sample of galaxies with companions which is an order of magnitude
larger than those used in previous studies.  Such a data set enables
us to investigate not only the effect of tidal interactions on star
formation within galaxies, but also to examine this as a function of
morphological type and nature of the perturbing galaxy. Interpretation
of our results is made easier by using a substantially complete
volume- and luminosity-limited sample.

We begin our discussion by briefly reviewing sources of possible error
and bias in our sample.  To estimate star-formation rate we have used
the \Halpha\ luminosity from the SDSS.  This was corrected for
extinction using measured \Hbeta data and then corrected for the
finite aperture size (see Section~\ref{sec:halpha-line}).  By
constraining the lower redshift limit of the sample to 0.03 the
aperture correction is minimised.  As shown in
\cite{2003ApJ...599..971H} this procedure leads to measurements of the
star-formation rate in agreement with other estimators, however, a
residual effect remains.  For systems in which star formation is
dominated by a nuclear starburst that falls within the SDSS fibre, the
aperture correction will tend to overestimate the star-formation rate
to some extent.  On the other hand, if the majority of star formation
is off-nuclear, or if the fibre is not accurately positioned on the
nucleus, then the aperture correction will not recover the true
star-formation rate.  Undoubtedly, these effects occur in measurements
of individual objects, however the data shown in
Figure~\ref{fig:apcoreffect} suggest that on the whole the aperture
correction is a useful way of estimating the global star-formation
rates.

We have also made use of {\it IRAS\/} FIR data to give an independent
estimate of star-formation rate for those objects in the volume
limited sample that are detected.  These systems are the most actively
star-forming systems in the sample, and therefore most likely to
suffer significant problems due to the aperture correction and
extinction.  We have found that the results based on \Halpha and FIR
are in good agreement suggesting that the \Halpha-based estimate of
the star-formation rate is sufficiently unbiased for the purposes of
this study. Further interpretation is made difficult by the coarse
resolution of \emph{IRAS;} if we normalise the FIR-derived SFR by just
the mass of the primary galaxy -- and ignore the possibility that the
companion also contributes to the flux -- it appears that \Halpha\
estimated SFR in the close pairs are approximately 30\% smaller than
those estimated using the FIR data: an indication perhaps that we
underestimate the extinction in these systems. An argument for
following this approach is that, at least for samples of bright
galaxies, \cite{2004AJ....127.3235S} and works referenced therein show
that in pairs of interacting galaxies only one is usually
infrared-luminous.  Alternatively, if we normalise the FIR-derived SFR
by the combined masses of the primary galaxy and companion, we find
that there is good agreement for the whole range of separations.

Although we have removed all systems which show any spectroscopic
evidence of having an AGN, it is possible that highly obscured AGN are
present in our sample.  Given our procedures, it seems likely that if
such systems are present their \Halpha\ fluxes will be dominated by
star formation. On the other hand, their FIR fluxes -- which more
closely measure bolometric output -- may be enhanced by the AGN; but,
this can only make the correlation between the \Halpha\ and FIR
measured star-formation rates worse.
 
As discussed in Section~\ref{sec:defin-catal-galaxy}, closest
companions were selected on the basis of \zband\ magnitude difference
in the projected separation range $3\times$ Petrosian half-light
radius $< r_{p} <300\,\unit{kpc}$, and using spectroscopic data in the
20\% of cases where it was available. This approach has the advantage
that our sample is complete to a minimum pair separation of about
15\,kpc and to a well defined \zband\ luminosity difference.  This
completeness comes at the cost of contamination by projection effect
pairs; we have shown that these only become significant at separations
greater than 150\,kpc. In these cases the distance to the closest
companion represents a lower-limit to the true separation.  Since this
contamination is only significant at larger separations it should not
affect the interpretation of our results.

The imposed minimum pair separation and intrinsic difficulty in
de-blending galaxies with overlapping disks does mean we will
inevitably miss very close pairs and merging systems.  We note that of
the 9 systems in the highest SFR bin
(Section~\ref{sec:res-pair-morphology}) that are classified as
actively merging, all bar one have been erroneously paired with
distant galaxies.

We observe an enhancement in the specific star-formation rate in
late-type systems with a close companion ($r_{p} \lesssim
50\,\unit{kpc}$).  This result is similar to that found by \cite{
2000ApJ...530..660B} and \cite{2003MNRAS.346.1189L}.  A quantitative
comparison to the results of \cite{2003MNRAS.346.1189L} is possible.
The functional form used to fit the data in \cite{2003MNRAS.346.1189L}
is a single power-law of the form $ \langle b \rangle =a_{1}
r_{p}^{\beta}$ where $b$ is the \emph{birth parameter\/} which should
be proportional to $\psi_{m}$.  We have fitted our data using a
similar functional form, restricting the range of projected
separations to that used by \cite{2003MNRAS.346.1189L} [$0<r_{p} <
100\times(0.7)^{-1}\,\unit{kpc}$].  After converting results to a
common distance scale we find that we obtain agreement with
\cite{2003MNRAS.346.1189L} only if we apply the additional constraint
$\psi_{m} > 2\times10^{-11}\,\unitp{yr}{-1}$, which accounts for a
bias towards more star forming systems in \cite{2003MNRAS.346.1189L}.

The larger sample size, and additional parameters available in the SDSS
have enabled us to examine these basic results in greater detail.

Although the strongest enhancement in the specific star-formation rate
occurs for small physical projected separations, we also find that for
late-type galaxies, there appears to be a systematic decrease of
specific star formation with increasing companion separation out to
300\,kpc.  A similar trend may also be present in the data of
\cite{2003MNRAS.346.1189L} (see their Figure~1) although
\citeauthor{2003MNRAS.346.1189L} do not reach this conclusion.  This
long-range correlation between star formation and pair separation is
seen only in late-type systems, i.e., those with concentration indices
$> 0.375$.  

It should be noted that the combination of significant projection pair
contamination at large separations and the morphological separation
into only three bins might contrive to make galaxies of slightly
earlier morphological type biased towards pairs at large separations
which could produce a correlation similar to that which is observed.

An interpretation of this effect as due to triggering of
star-formation would, however, be consistent with the predictions of
\cite{1996ApJ...464..641M}, who show that galaxies with weak bulge
components are more susceptible to formation of bars during
interaction.  The presence of the bar provides the mechanism for gas
flow towards the nucleus, leading to enhanced star formation over a
prolonged period of time, even when the galaxies are still widely
separated. Such weak-bulge systems constitute the highest
concentration index systems in our sample.  In contrast, systems with
strong bulges (i.e., lower concentration indices) are stabilised
against bar formation by the presence of the bulge. As a result, the
onset of star formation is delayed until the final stages of merger.

It is clear from our data that, if only actively star-forming systems
are considered (e.g., $\psi > 10\,\MSolar\,\unitp{yr}{-1}$,
Figure~\ref{fig:SFRbins}), the increase of the average star formation
at close separations is much higher. This suggests that the
significance of interactions will be somewhat overestimated in studies
which are biased towards selecting star-forming systems. Our data for
the FIR detected sub-sample demonstrate this quite clearly.

We also find a surprisingly tight relationship between the
concentration index and pair separation. The steep decline of the
concentration index at projected separations below 50\,kpc is most
easily explained if there exists a population of systems with nuclear
starbursts at these separations.  If this is the case, it is quite
likely that a few systems with strong disk components and intense star
formation have been misclassified, using the concentration index, as
intermediate types. It is possible such systems are responsible for
some, or even most, of the enhancement of SFR in galaxies of
intermediate type with close pairs.

The specific star-formation rate is found to decrease with increasing
recessional velocity separation out to the largest velocity
separations, suggesting that fly-by encounters, in which the two
systems are not on bound orbits, do contribute to enhanced star
formation. The specific star-formation rate appears to be a weaker
function of velocity difference than of projected physical separation,
however this may be a result of dilution of the velocity results due
to line of sight projection effects. Comparing our results with those
of \cite{2003MNRAS.346.1189L} we note a similar percentage drop in the
SSFR over the velocity range 0--200$\,\kms$, however, we do not
observe the marked change in gradient at $\sim 350\,\kms$ they
observe.

The nature of the perturber does not appear significant in determining
star formation enhancement.  We find no difference in the
effectiveness of triggering star formation between late-type /
late-type and late-type / early-type interactions.  Further, we find
no evidence for a dependence of the specific star-formation rate on
the mass of the companion.  This is in contrast both to the results of
\cite{1997ApJS..111..181D} and
\cite{2003MNRAS.346.1189L}. \cite{1997ApJS..111..181D} examine star
formation enhancement in a sample of 27 physical pairs consisting of a
main galaxy and a companion that has approximately half the
diameter. They find that the star formation enhancement is much
stronger ($\sim 50\%$ on average) in the minor components of the
pairs. In contrast, \cite{2003MNRAS.346.1189L}, dividing their pair
sample into minor and major merger candidates according to the
relative luminosity of the component galaxies, find that the bright
components of minor merger pairs show a higher probability of having
enhanced star formation. Comparing the star formation enhancement in
major mergers, they find no significant difference between the two
components.  Our results may be reconciled with those of
\cite{2003MNRAS.346.1189L}; as is clear from
Figure~\ref{fig:histmassdiff} the mean of the distribution for the
sub-sample with $ -2 <\Delta m_z <-1$ is greater than that for $0 <
\Delta m_z < 2$, however this is caused by small number statistics at
high specific star formation rates.

An interpretation consistent with our own results and those of
\cite{2003MNRAS.346.1189L} is as follows.  The probability of a galaxy
undergoing enhanced star formation depends strongly on the properties
of the galaxy itself (availability of gas etc.) and on the nature of
the orbital interaction with its companion.  The mass of the companion
appears to be of less importance -- at least up to the factor of $\sim
6$ maximum mass ratio that we probe.  There is a strong tendency for a
given system to be involved in an interaction with a lower-mass
companion (a `minor' interaction), but we believe this results simply
from the form of the luminosity function.  These results are still in
conflict with the findings of \cite{1997ApJS..111..181D} -- in this
case it is not clear how their sample selection and small number
statistics would effect their conclusions.

In this work, we have not explicitly considered the effect of
environment on galaxy star-formation rates. Our volume-limited sample
is selected from the MGS which only contains galaxies with
spectroscopic data, and therefore is biased slightly \emph{against}
galaxies in dense environments, where fibre collisions are more
likely. Our pair finding procedure however uses all the data, and
therefore does not suffer from this bias. Thus, although high-density
regions may be somewhat under-represented in our primary sample, the
pair separations should be unbiased.

As noted in the introduction, a number of studies have examined the
relationship between star-formation rates and the environment
\citep[e.g.,][]{2003ApJ...584..210G,2002MNRAS.334..673L,1998ApJ...499..589H}
-- these studies all find a reduction in the mean star-formation rate
in dense environments. However, interpreting these results is
difficult because of the well-known and very strong correlation
between morphological type and environment
\citep[e.g.,][]{1980ApJ...236..351D,2003MNRAS.346..601G}, and the correlation
of morphological type and star formation
\citep[e.g.,][]{1998ARA&A..36..189K}. Other galaxy properties are also known to
correlate with environment: galaxies are, for example, on average more
luminous in rich clusters
\citep[$M^{*}_{bJ}=-20.07\pm0.07$,][]{2003MNRAS.342..725D} and
marginally more luminous in groups
\citep[$M^{*}_{bJ}=-19.90\pm0.03$,][]{2002MNRAS.337.1441M} as compared
to the field
\citep[$M^{*}_{bJ}=-19.79\pm0.04$,][]{2002MNRAS.333..133M}.

Our results show enhanced specific star-formation rates at close pair
separations. The simplest interpretation, which we favour, is that
this is indicative of tidally-triggered star formation. Indeed,
tidally triggered conversion of gas into stars during cluster assembly
has been suggested as the \emph{cause} of the relation between density
of environment and star formation \citep[][and reference
therein]{2004MNRAS.348.1355B}. Alternatively, it could be a result of
some other process, also responsible for the observed correlation
between galaxy properties and environment. This however seems unlikely
for the following reasons.  If close pairs are more likely in dense
environments, the works cited above suggest that their star-formation
rate would, on average, be reduced and their luminosity increased
thereby further reducing their specific star-formation rate.  We,
however, observe the opposite effect. Our observations could be
explained by environmental effects if close pairs are in fact more
likely in the field where the average star-formation rate is higher. A
full analysis of these effects is beyond the scope of the present
paper, but will be the subject of a forthcoming investigation.

\section{Conclusions}

We have used SDSS DR1 imaging and DR2 spectroscopy products to
construct a catalogue of nearest companions for a volume and
luminosity limited sample of galaxies drawn from the Main Galaxy
Sample. Star formation rates for the volume-limited sample have been
calculated from extinction and aperture corrected \Halpha
luminosities, and {\it IRAS\/} data where available. We used \rband\
concentration indices as a proxy to morphological classification and
\zband\ luminosities to estimate galaxy masses. Our results indicate
that:
\begin{enumerate}
  \item The mean specific star formation rate of galaxies shows a
  clear (anti-)correlation with distance to the nearest companion.
  For early- and mixed-type systems this correlation is observed for
  projected separations below 50\,kpc; for late-type systems, it is
  observed out to 300\,kpc, the largest separation that we probe.
  \item We show this correlation is more pronounced in actively
  star-forming systems.
  \item We observe a decline in the mean star-formation rate with
  increasing recessional velocity difference for those galaxy pairs
  where a redshift has been measured for both galaxies. The magnitude
  of this effect is small when compared with projected separation.
  \item We observe a tight relationship between the concentration
  index and pair separation which peaks at $r_{p}=75\,\unit{kpc}$ and
  declines rapidly at smaller separations. We interpret this trend as
  due to the presence of tidally-triggered nuclear starbursts in systems
  with small projected separation.
 \end{enumerate}

\bibliographystyle{mn2e}
\bibliography{../main}

\section*{Acknowledgements}

    Funding for the creation and distribution of the SDSS Archive has
    been provided by the Alfred P. Sloan Foundation, the Participating
    Institutions, the National Aeronautics and Space Administration,
    the National Science Foundation, the U.S. Department of Energy,
    the Japanese Monbukagakusho, and the Max Planck Society. The SDSS
    Web site is http://www.sdss.org/.

    The SDSS is managed by the Astrophysical Research Consortium (ARC)
    for the Participating Institutions. The Participating Institutions
    are The University of Chicago, Fermilab, the Institute for
    Advanced Study, the Japan Participation Group, The Johns Hopkins
    University, Los Alamos National Laboratory, the
    Max-Planck-Institute for Astronomy (MPIA), the
    Max-Planck-Institute for Astrophysics (MPA), New Mexico State
    University, University of Pittsburgh, Princeton University, the
    United States Naval Observatory, and the University of Washington.

    This research has made use of the NASA/IPAC Extragalactic Database
    (NED) which is operated by the Jet Propulsion Laboratory,
    California Institute of Technology, under contract with the
    National Aeronautics and Space Administration.

    This research has made use of NASA's Astrophysics Data System.

    This research has made use of the NASA/ IPAC Infrared Science
    Archive, which is operated by the Jet Propulsion Laboratory,
    California Institute of Technology, under contract with the
    National Aeronautics and Space Administration.

    This publication makes use of data products from the Two Micron All
    Sky Survey, which is a joint project of the University of
    Massachusetts and the Infrared Processing and Analysis
    Center/California Institute of Technology, funded by the National
    Aeronautics and Space Administration and the National Science
    Foundation.

    We would like to thank Sebastian Jester for answering numerous
    queries about the SDSS. Finally, we thank the referee for detailed
    and helpful comments on the first version of the paper.

\label{lastpage}
\end{document}